\documentclass[12pt,preprint]{aastex}

\newcommand{\degree}{$^\circ$}
\newcommand{\Halpha}{H\ensuremath{\alpha}}
\newcommand{\Hbeta}{H\ensuremath{\beta}}
\newcommand{\Hgamma}{H\ensuremath{\gamma}}
\newcommand{\Hdelta}{H\ensuremath{\delta}}
\newcommand{\Heps}{H\ensuremath{\epsilon}}
\newcommand{\kms}{\ensuremath{\textrm{km~s}^{-1}}}
\newcommand{\ebmv}{E(\ensuremath{B}-\ensuremath{V})}
\newcommand{\gj}{SN~2005gj}
\newcommand{\ic}{SN~2002ic}
\newcommand{\tlike}{SN~1991T-like}
\newcommand{\snia}{SN~Ia}
\newcommand{\sneia}{SNe~Ia}
\newcommand{\snii}{SN~II}
\newcommand{\sniin}{SN~IIn}
\newcommand{\sneiin}{SNe~IIn}
\newcommand{\wl}{$\lambda$}
\newcommand{\is}{\emph{i}}
\newcommand{\vs}{{2,500~\kms}}
\newcommand{\vw}{{60~\kms}}
\newcommand{\lbol}{{$1.7 \times 10^{44}$~erg~s$^{-1}$}}

\slugcomment{To appear in Astrophysical Journal}

\begin{document}

\title{%
Nearby Supernova Factory Observations of SN 2005gj: Another
Type Ia Supernova in a Massive Circumstellar Envelope.
}

\author{%
 The Nearby Supernova Factory \\
 G.~Aldering,\altaffilmark{1}
 P.~Antilogus,\altaffilmark{3}
 S.~Bailey,\altaffilmark{1}
 C.~Baltay,\altaffilmark{8}
 A.~Bauer,\altaffilmark{8}
 N.~Blanc,\altaffilmark{2}
 S.~Bongard,\altaffilmark{1,5}
 Y.~Copin,\altaffilmark{2}
 E.~Gangler,\altaffilmark{2}
 S.~Gilles,\altaffilmark{3}
 R.~Kessler,\altaffilmark{7}
 D.~Kocevski,\altaffilmark{1,6}
 B.~C. Lee,\altaffilmark{1}
 S.~Loken,\altaffilmark{1}
 P.~Nugent,\altaffilmark{1}
 R.~Pain,\altaffilmark{3}
 E.~P\'econtal,\altaffilmark{4}
 R.~Pereira,\altaffilmark{3} 
 S.~Perlmutter,\altaffilmark{1,6}
 D.~Rabinowitz,\altaffilmark{8}
 G.~Rigaudier,\altaffilmark{4}
 R.~Scalzo,\altaffilmark{1}
 G.~Smadja,\altaffilmark{2}
 R.~C. Thomas,\altaffilmark{1}
 L.~Wang,\altaffilmark{1}
 B.~A.~Weaver\altaffilmark{1,5}
}


\altaffiltext{1}{Physics Division, Lawrence Berkeley National Laboratory, 
1 Cyclotron Road, Berkeley, CA, 94720}
\altaffiltext{2}{Institut de Physique Nucl\'eaire de Lyon, UMR5822,
    CNRS-IN2P3; Universit\'e Claude Bernard Lyon~1, F-69622 Villeurbanne
    France}
\altaffiltext{3}{Laboratoire de Physique Nucl\'eaire et des Hautes Energies
    IN2P3 - CNRS - Universit\'es Paris VI et Paris VII, 4 place Jussieu
    Tour 33 - Rez de chauss\'ee 75252 Paris Cedex 05}
\altaffiltext{4}{Centre de Recherche Astronomique de Lyon,
    9, av. Charles Andr\'e, 69561 Saint Genis Laval Cedex}
\altaffiltext{5}{University of California, Space Sciences Laboratory, Berkeley, CA 94720-7450}
\altaffiltext{6}{Department of Physics, University of California, Berkeley,
CA 94720}
\altaffiltext{7}{Kavli Institute for Cosmological Physics,
    The University of Chicago, Chicago, IL 60637}
\altaffiltext{8}{Department of Physics, Yale University, New Haven, CT 06250-8121}

\begin{abstract}

We report the independent discovery and follow-up observations of
supernova 2005gj by the Nearby Supernova Factory. This is the
second confirmed case of a ``hybrid'' Type Ia/IIn supernova, which
like the prototype \ic, we interpret as the explosion of a white
dwarf interacting with a circumstellar medium. Our early-phase photometry of
\gj\ shows that the strength of the interaction between the supernova
ejecta and circumstellar material is much stronger than for \ic. Our
first spectrum shows a hot continuum with broad and narrow \Halpha\ emission.
Later spectra, spanning over 4 months from outburst, show clear
Type~Ia features combined with broad and narrow \Hgamma, \Hbeta, \Halpha\
and \ion{He}{1}~\wl\wl5876,7065 in emission.  At higher resolution,
P~Cygni profiles are apparent. Surprisingly, we also observe an inverted
P~Cygni profile for [\ion{O}{3}]~\wl5007. 
We find that the lightcurve and measured velocity of the unshocked
circumstellar material imply mass loss as recently as 8 years ago.
This is in contrast to \ic, for which an inner cavity in the
circumstellar material was inferred.  Within the context of the
thin-shell approximation, the early lightcurve is well-described by a
flat radial density profile for the circumstellar material. However, our
decomposition of the spectra into Type~Ia and shock emission components
allows for little obscuration of the supernova, suggesting an aspherical
or clumpy distribution for the circumstellar material. We suggest that
the emission line velocity profiles arise from electron scattering rather
than the kinematics of the shock. This is supported by the inferred high
densities, and the lack of evidence for evolution in the line widths.
Ground- and space-based photometry, and Keck spectroscopy, of the host
galaxy are used to ascertain that the host galaxy has low metallicity
($Z/Z_{\odot}<0.3$; 95\% confidence) and that this galaxy is undergoing
a significant star formation event that began roughly $200\pm70$~Myr ago.
We discuss the implications of these observations for progenitor models
and cosmology using Type~Ia supernovae.

\end{abstract}

\keywords
{
galaxies: abundances ---
stars: winds, outflows --- 
supernovae: general --- 
supernovae: individual (SN~2005gj, SN~2002ic)
}

\section{Introduction}

Type~Ia supernovae (\sneia) are very important in
astrophysics. As standardized candles they played the leading
role in the determination that the mass-density of the Universe is
sub-critical and that the expansion of the Universe is accelerating
\citep{perl98,garnavich98,riess98,perl99}.  They continue to play a major
role in on-going efforts to determine the cause of this acceleration
\citep{knop03,tonry03,riess04,astier06} and in refining the determination
of the Hubble constant \citep{riess05}. \sneia\ are a major source
of chemical enrichment in galaxies, and may play a major role in the
development of low-mass galaxies \citep[e.g.][]{ferrara00}.  For these
many reasons, a long-standing goal has been to precisely identify the
progenitor systems of \sneia. Of special interest for cosmology --- where
high-redshift \sneia\ are compared to low-redshift \sneia\ to deduce
cosmological parameters --- is the question of whether or not \sneia\
may arise from multiple channels \citep[e.g.,][]{howell01,scannapieco05},
acting on significantly different timescales, and in particular whether
the relative contributions of multiple channels could act in such a way
as to lead to biased cosmology results \citep{drell00,leib01}.

Historically speaking, the two most strongly favored \snia\ progenitor
scenarios both involve the thermonuclear disruption of white dwarf stars
\citep[for an overview, see][]{bra95}.  In the \emph{single degenerate}
model, a carbon-oxygen white dwarf accretes matter shed by a
less-evolved companion star until it reaches the Chandrasekhar mass and
explodes.  The \emph{double degenerate} scenario consists of the
coalescence of a binary white dwarf system.  The role of mass loss in
the former scenario suggests an observational test to discriminate
between the two models.  The unambiguous detection of circumstellar
hydrogen in the spectrum of a \snia\ would be interpreted as support for
the single degenerate scenario.  The search for this signature has
resulted in mostly nondetections and upper limits on mass loss from the
progenitor system, until recently.

The spectra of the peculiar \snia~2002ic \citep{woo02,ham03} were the first to clearly
reveal the interaction of the ejecta of one of these events with its
circumstellar medium (CSM).  The spectra of \ic\ resembled those of the
spectroscopically peculiar, overluminous \snia~1991T, with features
characteristic of a \snia\ though ``diluted'' \citep{ham03}.  Most
remarkably, these spectra exhibit narrow \Halpha\ emission consistent
with the redshift of the SN \citep[$z = 0.0667$,][]{kotak04} strongly
reminiscent of \sneiin\ \citep[``n'' for narrow hydrogen
emission,][]{sch90}.  More circumstantial evidence obtained by comparing
late time spectra of \sneiin\ 1997cy and 1999E with \ic\ \citep{ham03,deng04}
suggests that some fraction of \sneiin, ostensibly core-collapse SNe,
may in fact be \sneia\ veiled by a substantial CSM.

Various interpretations of the observations of \ic\ have appeared in the
literature.  Motivated by the large mass inferred for the CSM,
\citet{ham03} suggested that the progenitor system contained a massive
asymptotic giant branch (AGB) star which shed the material prior to
explosion.  The AGB star could be a binary companion star, or the SN
progenitor itself \citep[a Type 1.5 SN,][]{ibe83}.  Most analyses agree
that the density of the CSM was high, that a few tenths to several solar
masses of material were shed, and that the mass loss rate was high or
the CSM was clumpy \citep{kotak04,chugai04c,wang04}.  Spectropolarimetry
suggested that the geometry of the CSM was aspherical, and perhaps
equatorially condensed \citep{wang04,uenishi04}.  A light curve
analysis \citep{woo04} implied a delay in the onset of the circumstellar
interaction, suggesting that the mass loss phase had ended or trailed
off in the years preceding the SN or that the inner CSM had been 
cleared by a fast wind, both consistent with the formation of a
proto-planetary nebula (PPN; \citet{kwok93}) by an AGB star.

Instead of an AGB scenario, \citet{livio03} suggested that \ic\
was in fact the product of a double degenerate merger during the common
envelope (CE) phase of a binary; the white dwarf spirals in to coalesce
with its companion's core.  However, the disk-like geometry and large
radial extent of the CSM inferred from spectropolarimetry \citep{wang04}
and coalescence timescales that are incompatible with the timescale
inferred for the mass ejection \citep{chu04b} seem to rule out the
CE coalescence model for \ic.

Motivated in part by the inferred presence of considerable CSM
surrounding \ic, \citet{wang05a} advanced a hypothesis to explain the
low values $R_B \equiv A_B /$\ebmv\ often derived from observations of
\sneia. Instead of the value $R_B\sim4.1$ expected for dust extinction
in the Galaxy and LMC, values $R_B\sim2$--3 are often found
\citep{tripp99,phillips99,wang03,knop03,wang05b}.  In this hypothesis,
dust in the immediate vicinity of the SN or at the highest velocities of
its ejecta scatter SN light into the observer line of sight, thus adding
back light and effectively reducing derived values of $R_\lambda$ in the
optical.  The source of the dust may be a PN or PPN formed during the
evolution of the progenitor system.  This explanation suggests a
continuum of models for ejecta-CSM interaction in \sneia.  \sneia\ with
more normal values of $R_B$ could represent cases where the PN has had
enough time to disperse into the host interstellar medium.  Cases where
$R_B$ is low but there is no CSM interaction signature could occur in
diffuse PNe.  Still others (such as \ic) with a strong CSM interaction
signature might originate in more compact PNe or PPNe.

Here we report on our observations of the recent \gj, which has much in
common with \ic.  \citet{pri05} observed the development of spectral
features characteristic of a \snia, during their follow-up of the
discovery of \gj\ \citep{cbet247,fri05}, in what had originally appeared
to be a \sniin.  The Nearby Supernova Factory \citep[SNfactory]{ald02}
had independently discovered \gj\ in images taken 2005 September~29,
and acquired optical spectroscopy on 2005 October~3 which showed the
strong, comparatively narrow Balmer \Halpha\ line that is the defining
characteristic of the IIn class, and no visible \snia\ features.  Both
\gj\ and \ic\ were discovered by ``blind'' SN surveys (conducted without
targeting specific galaxies) --- \gj\ by the SDSS and the SNfactory,
and \ic\ by the SNfactory prototype search \citep{woo02}.  The two SNe
are hosted by low-luminosity galaxies at about the same redshift, and
their spectra at about 70 days after explosion are remarkably similar.

There are, however, important differences between the two events.  Our
earliest spectrum of \gj\ indicates a more intense, earlier interaction
with the CSM than in the case of \ic.  Additionally, the light curve of
\gj\ is brighter than that of \ic, supporting the conclusion that the
\gj\ ejecta-CSM interaction is stronger.  Still, X-ray \citep{imm05}
and radio \citep{sod05} observations resulted in non-detections, as was
the case for radio observations of \ic\ \citep{berger03,stockdale03}.

The organization of this paper is as follows.  In \S2 we present
new photometry and spectroscopy of \gj\ covering numerous epochs
extending more than four months past the explosion date. We also describe
spectroscopic observations of the host galaxy. In \S3 we analyze the \gj\
observations in the context of a two-component SN plus CSM interaction
model in order to place constraints on the amount of CSM and the relative
contributions of SN and CSM light to the observations.  \S4 presents
an analysis of the host galaxy spectroscopy and photometry. These are
brought together in \S5 in a discussion of our results and their possible
implications for SN progenitor systems and cosmology.  \S6 summarizes
our findings. We take $H_0 = 70$ km s$^{-1}$ Mpc$^{-1}$ throughout
this paper.

\section{Data}

\subsection{Photometry}

\subsubsection{Search, Discovery, and Follow-up}

The SNfactory independently discovered \gj\ (our SNF20050929-005) on
2005 September 29.45 UCT (JD 2453642.95) in wide-field images obtained
using the QUEST-II CCD camera \citep{rab03} on
the Samuel Oschin 1.2m telescope on Mount Palomar, CA, in collaboration
with the JPL Near-Earth Asteroid Tracking (NEAT) component of the
Palomar-QUEST Consortium.  These asteroid and SN
search images consist of 60 second exposures in an RG610 filter.  Each
field is visited three times within an hour in order to detect the
motion of asteroids (which we reject) and eliminate cosmetic defects.
Subsequent to discovery, additional images of \gj\ were obtained by
NEAT, as well as by the QUEST collaboration. QUEST images were obtained
in drift-scan mode, using either Johnson \emph{U}, \emph{B}, \emph{R},
\emph{I} filters or Gunn \emph{r}, \emph{i}, \emph{z} filters.

The images are transfered from Mount Palomar to the High Performance
Storage System (HPSS) at the National Energy Research Scientific
Computing Center (NERSC) in Oakland, CA via the wireless HPWREN
network\footnote{Information on HPWREN (PI: H.-W. Braun) is available at
http://hpwren.ucsd.edu.} and the ESnet network.\footnote{Information on
the US Department of Energy's ESnet is available at http://www.es.net.}
These images are processed using the Parallel Distributed System
Facility (PDSF) at NERSC.  The first level of processing decompresses
the images, subtracts an average bias image, and divides by a normalized
median sky flat image constructed from images taken around the same time
on the same night.  An object-finding algorithm identifies objects that
have counterparts in the USNO-A1.0 POSS-E catalog \citep{mon96}, and
these matches are used to determine the image astrometry.  

Images of a given field taken on the same night are spatially aligned,
convolved to a point spread function (PSF) common to these images and
their reference images, scaled in flux, and coadded to form a new image.
For NEAT images, archived unfiltered NEAT images from 2001-2003 are
used as references.  For QUEST drift-scan images, references are taken
from QUEST drift-scan runs in 2003 and 2004.  Reference images covering
the same field are astrometrically aligned to match the new images, PSF
convolved, then coadded.  This reference sum is then scaled and subtracted
from the new images. Residual sources passing a set of image parameter
cuts are automatically identified and passed on to human scanners for
final confirmation.

\subsubsection{Photometry and Lightcurve}
\label{sec:photometry}

For consistency in the analysis, the NEAT data were reprocessed using
the same 5 reference images for all new images.  For each night, the
photometry of one of the new images was calibrated against SDSS
\is\ magnitudes \citep{sdssdr4} for stars on that image. Between 90 and
199 stars were used, depending upon the observing conditions each
night.  Aperture photometry for each non-saturated star on the image
was compared to the SDSS \is\ magnitude and an image zeropoint was fit,
including (\emph{r}-\is) and (\emph{i}-\emph{z}) color corrections to
account for the differences between the RG610, Johnson \emph{I}, or
Gunn \emph{i}, and SDSS \is\ filters.  All other images used in the
subtraction were photometrically calibrated against this new image.

Aperture photometry was performed on \gj\ using these zeropoints to
produce magnitudes calibrated to the SDSS \is\ band.  The color of
\gj\ was interpolated from QUEST drift-scan images taken in several
filters, or from our spectrophotometry (see \S2.2).
Additionally, flux calibrated SNIFS spectra, described in \S\ref{sec:SNIFS},
were used to synthesize \is\ magnitudes and
(\emph{r}-\is) and (\is-\emph{z}) colors on
the SDSS system for dates JD-2453000 of 647.09, 699.96, 706.96, and
706.98.

\gj\ was not evident on images taken (under poor weather conditions)
on 2005 September 23.4. {\sc IRAF}\footnote{{\sc IRAF} is distributed
by the National Optical Astronomy Observatories, which are operated by
the Association of Universities for Research in Astronomy, Inc., under
cooperative agreement with the National Science Foundation.} aperture
photometry of the subtracted image at the location of \gj\ was used
to establish that a 2$\sigma$ upward fluctuation of the background would
have corresponded to an object having \is = 20.15. Therefore we quote this as the
limiting magnitude for \gj\ on that day.

Observing conditions on 2005 November~18.4 were cloudy and there were
an insufficient number of stars on the image to perform the necessary
astrometric and PSF matching needed for our automated subtraction
pipeline.  Instead, these images were individually processed with {\sc
IRAF} using 5 nearby stars to determine the magnitude zeropoint.  The
errors on the 2005 November~18 photometry measurements are dominated by
the large sky background.

To compare with \ic\ \emph{V}-band data and models, \gj\ \emph{V}-band
magnitudes were estimated using extrapolations from Johnson \emph{B},
Johnson \emph{R}, Gunn \emph{r}, and RG610 filters, using the shape
of the SNIFS spectra for color corrections.  Nights on which we have
QUEST data in multiple filters confirms that this extrapolation method
produces photometry results consistent within the stated uncertainties.
When available, SNIFS spectra were also used to synthesize \emph{V}-band
photometry.

\begin{figure}
 \centering
 \includegraphics[width=0.750\textwidth,clip=true]{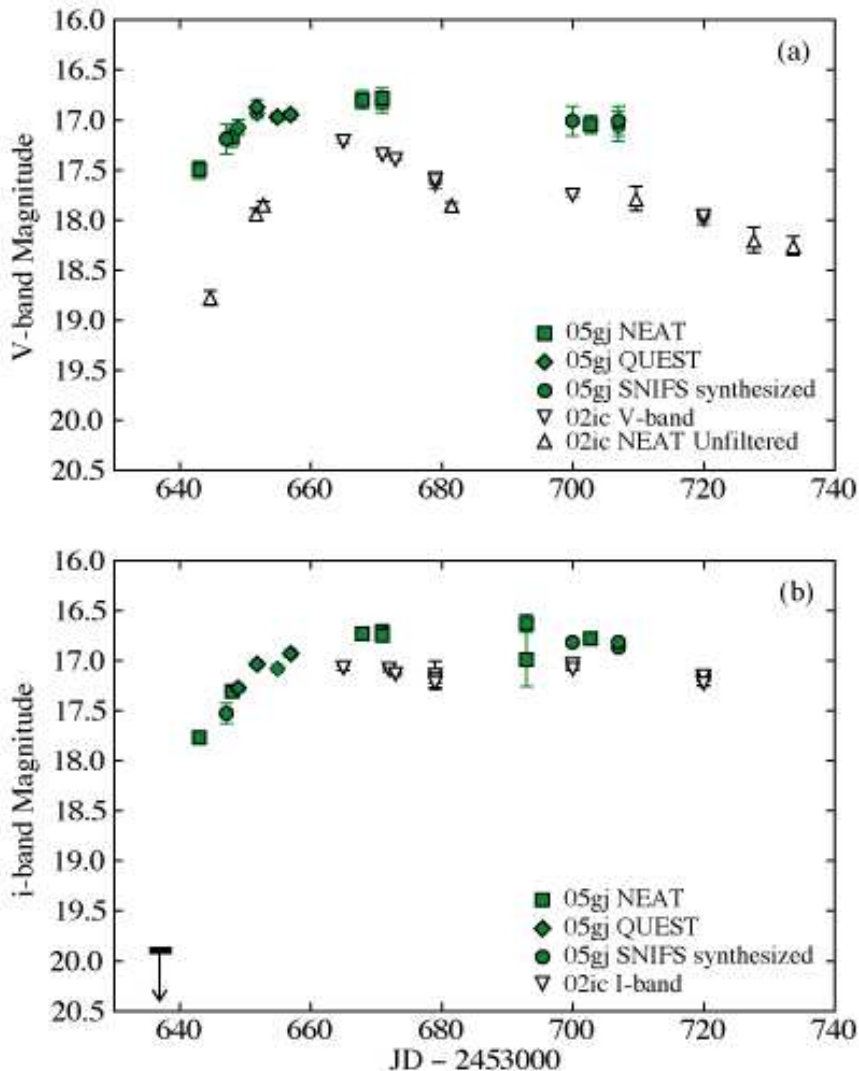}
 \caption{Lightcurve of \gj\ in \emph{V}-band (a) and \emph{i}-band (b).
   \ic\ observations are overlaid in \emph{V}-band and NEAT unfiltered (a)
   and \emph{I}-band (b).  The \gj\ \emph{V}-band data are extrapolations
   from other filters based upon the shape of the SNIFS spectra.
   The time axis of the \ic\ data has been
   shifted to align the estimated
   explosion dates of 2002 November~5 and 2005 September~22.6 UTC.
   All datapoints have been extinction corrected, and
   the \ic\ datapoints have been adjusted for the cosmological effects ---
   but not the bandpass shift --- due to the small difference in redshift.
   \cite{woo04} report that the 2002---2003 NEAT unfiltered photometry
   is comparable to \emph{V}-band photometry to within $\pm 0.05$ mag.
 }
 \label{fig:lightcurve}
\end{figure}

Table~\ref{tab:photometry} lists the measured \is\ photometry and
Table~\ref{tab:photometry_V} lists the derived \emph{V} photometry.
Correction for Galactic extinction, which in the direction of \gj\ is
\ebmv\ $= 0.121$ ($A_B = 0.52$~mag) according to the maps of \citet{sfd},
has not been applied in these tables.  Figure~\ref{fig:lightcurve}
illustrates our lightcurve data, including Galactic extinction
corrections.  Photometry points for \ic\ are overplotted for comparison
\citep{ham03, woo04}.  The \ic\ points have been extinction corrected and
adjusted to account for cosmological effects arising from the (small)
difference in redshift (0.0667 vs. 0.0616, see \S\ref{sec:Keck}). No
adjustment for redshift-dependent bandpass differences has been made,
since this effect is negligible for these filters and this redshift
difference.

The assigned magnitude uncertainties are predominantly from the
dispersion measured for the zeropoint calibration, which includes
statistical uncertainties arising from Poisson fluctuations as well as
residual errors from flat fielding, fringing, etc.  The QUEST images
taken with the Johnson \emph{I} filter also had significant color
correction terms (0.250 and 0.285), which combined with the color
uncertainty of \gj\ for those dates contributed several hundredths of a
magnitude to the uncertainty.

A quadratic fit to the flux on the first 40 days implies an explosion
date of JD~2453636.1 (2005 September~22.6~UTC), slightly before our
upper limit of \is\ $< 20.15$ on September~23.4.  In \S \ref{sec:lcfits}
we will use a CSM interaction model to refine this estimate.

\subsection{Optical Spectroscopy}

Our discovery of \gj\ on 2005 September~29.45~UTC prompted our follow-up
confirmation spectroscopy on 2005 October~3.59~UTC (JD 2453647.09).
We obtained additional optical spectroscopy at multiple epochs spanning
four months after explosion. Here we detail the spectroscopic reductions
and describe the major features observed. Table~\ref{tab:spectra}
provides technical details of these observations.

\subsubsection{SNIFS Spectroscopy}
\label{sec:SNIFS}

Observations of \gj\ were obtained with the SuperNova Integral Field
Spectrograph \citep[SNIFS,][]{ald02,lantz04}, operated by the SNfactory.
SNIFS is a fully integrated instrument optimized for automated observation
of point sources on a diffuse background over the full optical window
at moderate spectral resolution.  It consists of a high-throughput
wide-band pure-lenslet integral field spectrograph \citep[IFS, ``\`a
la TIGER,''][]{1995A&AS..113..347B, 2000ASPC..195..173B, bacon01}, a
multi-filter photometric channel to image the field surrounding the IFS
for atmospheric transmission monitoring simultaneous with spectroscopy,
and an acquisition/guiding channel.  The IFU possesses a fully filled
$6'' \times 6''$ spectroscopic field-of-view subdivided into a grid of
$15 \times 15$ spatial elements (spaxels), a dual-channel spectrograph
covering 3200--5200 \AA\ and 5100-11000 \AA\ simultaneously, and an
internal calibration unit (continuum and arc lamps).  SNIFS is permanently
mounted on the south bent-Cassegrain port of the University of Hawaii
2.2-m telescope (Mauna Kea), and is operated remotely.  The SNIFS spectra
of \gj\ were reduced using our dedicated data-reduction procedure,
similar to that presented in \S~4 of \citet{bacon01}. Here we briefly
outline the process.

After standard CCD preprocessing and subtraction of a low-amplitude
diffuse-light component, the 225 spectra from the individual lenslets of
each SNIFS exposure were extracted from each blue and red spectrograph
exposure, and re-packed into two $(x,y,\lambda)$-datacubes. This highly
specific extraction is based upon a detailed optical model of the
instrument including interspectrum crosstalk corrections.  The datacubes
were then wavelength-calibrated, using arc lamp exposures acquired
immediately after the science exposures, and spectro-spatially
flat-fielded, using continuum lamp exposures obtained during the same
night. Cosmic rays were detected and corrected using a 3D-filtering
scheme applied to the datacubes.

SN and standard star spectra were extracted from the 225 spaxels, $s$, of
each $(x,y,\lambda)$-datacube using a wavelength-dependent Gauss-Newton
PSF fit method by minimizing $\chi^2$.  The
adopted PSF model is
\begin{equation}
{\rm model}(s, \lambda) = S_{\rm SN} (\lambda) \cdot
I_{\rm PSF}(s, \vec x_{\rm SN}(\lambda_{\rm ref}), \lambda, \dots) + C(\lambda)
\end{equation}
where $S_{\rm SN}$ is the supernova spectrum, $I_{\rm PSF}$ is the PSF
integrated over one spatial element, and $C(\lambda)$ is a spatially
uniform background distribution.  Free parameters include the position
of the SN at an arbitrary reference wavelength \citep[the position at
other wavelengths follows an atmospheric differential refraction function;
see][]{fil82}, the position angle of the microlens array on the sky,
the SN spectrum, the background distribution, and the input parameters
for the chosen PSF (seven in our case, not including the SN position).
Uncertainties used in the $\chi^2$ minimization incorporate the detector
gain and readout noise in the calculation of the detector and photon
noise.

The PSF is modeled as a weighted sum of two bidimensional Gaussians
designed to account for variation near both the core and wings.  The FWHM
of each Gaussian is wavelength dependent, going as $\lambda^\gamma$.
Fit parameters include the orientation angle of the elliptical Gaussians,
the FWHMs at one arbitrary wavelength for each Gaussian, two $\gamma$
values, and a relative scaling between the Gaussians.

An attenuation estimate for each exposure was made using the stars
observed by the SNIFS multifilter photometric channel.  Objects in
the field (spatially subdivided into five regions each monitoring a different
wavelength range) were detected using SExtractor \citep{bertin96}, and their
fluxes measured using an adapted version of the Supernova Legacy Survey
aperture photometry code, Poloka.  For the objects that were matched
over all nights, the fluxes in each filter were summed, giving a total
flux per filter per night.  These fluxes were then normalized by the
flux on a specific night considered photometric, allowing us to make
an estimation of the relative flux extinction in each filter band for
the observations on non-photometric nights.  Seeing and the stability
of the atmospheric transmission were assessed using SNIFS guider video
frames acquired during our exposures.

Variable cloud conditions prevailed on the night of 2005 October~3~UTC.
The multi-filter transmission measurement indicates an average (gray)
correction factor of $3.74 \pm 0.09$.  Conditions on the night
of November~25 were photometric.  The night of December~2 was not
photometric, however, an average transmission correction of $1.01 \pm
0.01$ was derived from the multifilter, indicating that conditions were
mostly clear at the time \gj\ was observed.  These correction factors
are applied to the spectra when synthesizing magnitudes and measuring
line luminosities.

\begin{figure}
 \centering
 \includegraphics[width=0.725\textwidth,clip=true]{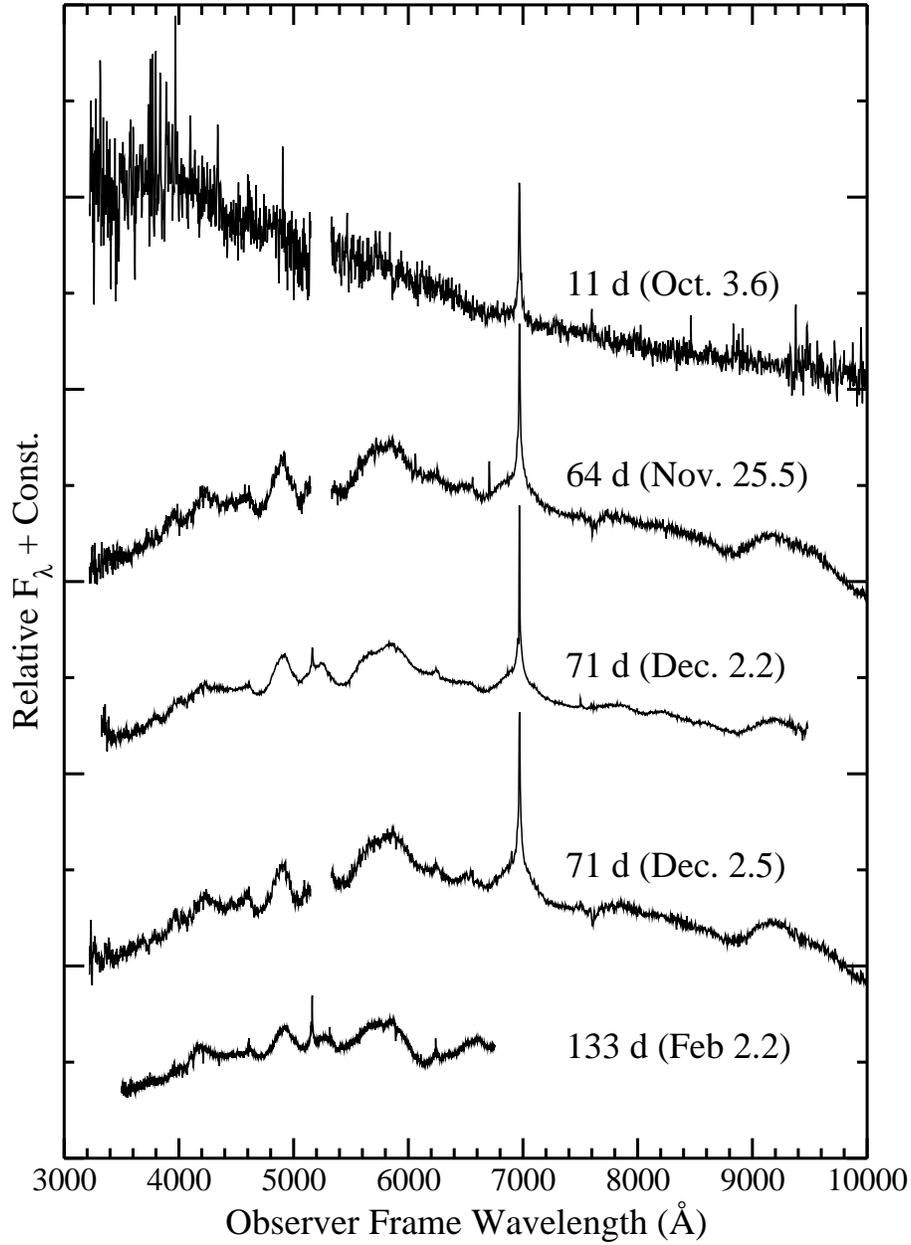}
 \caption{Combined spectroscopy of \gj.  The top two spectra were
 collected with SNIFS, the third with LRIS, the fourth again with SNIFS
 and the fifth with LRIS on the dates shown.  The SNIFS spectrum from
 December~2.5 is a coaddition of two spectra from the same night.  The
 age of each spectrum with respect to an assumed explosion date of 2005
 September~22.6~UTC is shown.}
 \label{fig:spectra}
\end{figure}

The final flux-calibrated spectra appear in Figure~\ref{fig:spectra}.
From the top, the first, second and fourth spectra were acquired with
SNIFS.  The wavelength range displayed corresponds to the region where
we are most confident in the calibration and extraction.  In particular,
the region surrounding the dichroic has been removed for clarity, as
the signal-to-noise here is low and is not always properly treated in
the current reduction pipeline.  The flux solution is conservatively
estimated to be accurate to within 5\% on November 25 and December~2,
but perhaps only 10-15\% on October~3.  These limits are imposed by the
simplified flux calibration procedure used here, in which no attempt has
been made to derive an independent atmospheric extinction curve for the
nights observed, but where a standard Mauna Kea atmospheric extinction
curve \citep{beland88} has been adopted. A full calibration solution
for each night is still in development.

Initial spectroscopy was obtained 4 days after our discovery.  As
Figure~\ref{fig:spectra} shows, at this early phase \gj\ resembled a
\sniin\ (a featureless continuum punctuated by a strong, narrow \Halpha\
emission feature), rather than a \snia\ that the SNfactory normally
would have continued to follow.  Later SNIFS spectra reveal the
continued presence of the \Halpha\ feature.  In the day~64 spectrum we
also detect a weak emission feature due to
\ion{He}{1} \wl 5876. On day~71 we also detect \ion{He}{1} \wl
7065.  Broad features, attributable to an underlying SN at about a month
after maximum, include \ion{Ca}{2} H\&K and infrared triplet, and
absorption from \ion{Fe}{3} at 4700 \AA\ and 5400 \AA.  In these spectra
there is no detectable \ion{Na}{1}~D absorption to indicate reddening by
the host galaxy.

\subsubsection{Keck Spectroscopy}
\label{sec:Keck}

We first observed \gj\ on the Keck~I telescope on 2005 December~2.3~UTC
using the Low Resolution Imaging Spectrograph \citep[LRIS][]{lris}.
The spectrograph was configured with the 5000~\AA\ blaze, 300~l/mm grism
in the blue channel, and the 8500~\AA\ blaze, 400~l/mm grating in the
red channel. A dichroic with crossover at 6800~\AA\ was used to separate
the light into red and blue channels, as the primary science that night
targeted high-redshift SNe.  Conditions were non-photometric during
this part of the night --- prior to the clear conditions that prevailed
later when our SNIFS spectrum was obtained --- and seeing varied between
$0\farcs7$ and $1\farcs0$.  A 900~sec integration using a 1\arcsec\ wide slit was
obtained, with a position angle of 0\degree\ in order to pass through both
the SN and its apparent host galaxy. Data were taken at airmass 1.08,
making corrections for atmospheric differential refraction negligible.

The data were overscan-subtracted, bias-subtracted, flat-fielded using
internal lamp exposures, wavelength-calibrated, and extracted using
standard {\sc IRAF} procedures. The wavelength calibration arcs were
obtained at the beginning the night; since LRIS suffers from flexure,
the wavelength zeropoints were adjusted for each spectrum using
the night sky lines referenced to the high resolution night sky spectrum
of \citet{uves_sky}. Flux calibration and telluric feature removal
was performed using the standard stars BD+174708, Feige~34, Feige~67,
Feige~110, and HZ~44.  For blue standard stars, the 6000-6800~\AA\
region of the blue channel spectrum shows contamination by second-order
light. This was accounted for when determining the flux correction, and
we find good agreement in the shape of the \gj\ spectrum between the blue
and red channels over the 6620--6770~\AA\ region where they overlap. A
standard Mauna Kea atmospheric extinction curve \citep{beland88}
was assumed, since non-photometric conditions precluded a direct
determination of the extinction. The resulting spectrum of \gj\ is
shown in Figure~\ref{fig:spectra}.  The S/N ratio of the spectrum is 100
or greater per 1.43~\AA\ pixel over the range 4000-9000~\AA.

While correction for atmospheric differential refraction is unimportant
for this spectrum, since atmospheric seeing is wavelength dependent 
the amount of light lost at the LRIS slit also depends on wavelength.
This effect is important for slit-widths comparable to the seeing
--- as with these observations --- and its incorporation into the
flux calibration is very dependent on the seeing and mis-centering
experienced by the standard stars relative to that experienced by
the target. As SNIFS does not suffer from this effect, we defer to
the SNIFS spectrum taken this same night when performing fits to the
flux over a long wavelength baseline.

The full 2D LRIS spectrum shows narrow \Halpha\ extending along the
major axis of the host galaxy. From this narrow emission we measure a
host galaxy heliocentric redshift of $0.0616 \pm 0.0002$. The peak of the
narrow emission from \gj\ is redshifted with respect to the host by $90
\pm 50$~km/s.  In addition to the spectrum of \gj, a region of the host
galaxy located $3\farcs4$ south of \gj, showing strong \Halpha\ emission
and relatively free from contamination by \gj, was extracted from the
LRIS red channel. This spectrum is analyzed in \S\ref{sec:spectral_analysis}

As in the SNIFS spectrum, \Halpha, \ion{He}{1} \wl 5876, and \ion{He}{1}
\wl 7065 are visible in emission from \gj.  In addition, \Hbeta\ emission
is clearly seen in the LRIS spectrum, and has a shape consistent with
that seen in \Halpha. \Hgamma\ is also detected, but \Hdelta\ is not.
With the higher resolution used for the LRIS spectrum, P~Cygni profiles
are apparent in the \Halpha\ and \Hbeta\ lines.  \ion{He}{1} \wl 4471 and
\ion{He}{1} \wl 6678 are not detected either in emission or absorption.
Weak (equivalent width EW$\sim$0.4--0.7~\AA), unresolved H9, H8,
and \ion{Ca}{2} H\&K/\Heps\ are detected in absorption. \ion{Na}{1}~D
absorption from the Galaxy is detected, having EW~=~0.6~\AA\ from the sum
of the (unresolved) D$_1$ and D$_2$ components; \ion{Na}{1}~D absorption
at the host galaxy redshift is not detected in this spectrum. Rather, at
this location there appears to be either a flat-topped high-velocity red
tail associated with \ion{He}{1} \wl 5876 or else a complex combination of
\ion{Na}{1}~D emission plus absorption, similar to that seen for SN~1994W
by \citet{chugai04a}. Also of interest is an inverted P~Cygni profile in
[\ion{O}{3}]~\wl 5007.

\gj\ was observed a second time with LRIS, on 2006 February~2.2~UTC,
using the 4000~\AA\ blaze, 600~l/mm grism in the blue and the 7500~\AA\
blaze, 1200~l/mm grating in the blue. This set-up allowed us to obtain the
higher resolution required to better examine the shapes of the \Hbeta\
and \ion{He}{1} \wl 5876 emission lines and higher level Balmer absorption
lines, and to search for \ion{Na}{1}~D absorption indicative of host-galaxy
extinction. The night was clear, and the seeing was $0\farcs9$ for
this 900~sec exposure. Processing followed the steps described for
2005~December~2, except that in this case the flux calibration was
obtained using G191B2B.

The blue Keck LRIS spectrum is remarkably similar to the spectrum taken two
months earlier.  The Balmer series from \Hdelta\ up to the Balmer limit
is clearly seen in absorption in this spectrum. \Hbeta\ and
\Hgamma\ show clear P~Cygni profiles, while \Hdelta, \Heps, and H8 show
strong absorption but only a hint of emission. The absorption of the
strong Balmer absorption lines and \ion{Ca}{2}~H, appear asymmetric,
with enhanced wings on the blue side.  The inverted P~Cygni profile for
[\ion{O}{3}]~5007\AA\ is also much more clear.  The blue spectrum also
exhibits narrow [\ion{O}{2}]~\wl\wl 3727,3729, most likely from the
host galaxy. These lines are shown in Figure~\ref{fig:linezoom}.

In the higher resolution red spectrum \ion{He}{1} \wl
5876 clearly exhibits a P~Cygni profile.  This is modeled in
\S\ref{sec:spectral_analysis}. In addition, very little \ion{Na}{1}~D
absorption from the host can be detected. The derived absorption
is EW(D1)~=~0.16~\AA\ and EW(D2)~=~0.12~\AA. These values are
sensitive to the setting of the ``continuum'' level, which in
this case is comprised of the broad component of \ion{He}{1} \wl
5876 (see \S\ref{sec:spectral_analysis}). We note that these lines
require a slightly smaller redshift, z=0.0613, within the uncertainty
quoted for the redshift determination from December~2.  However,
it is possible that these lines arise from material approaching by
$\sim90$~\kms. There is some hint that these weak absorption features
could be due to P~Cygni profiles associated with \ion{Na}{1}~D from
the CSM. Galactic \ion{Na}{1}~D is also detected quite clearly. We
measure EW(D1)~=~0.41~\AA\ and EW(D2)~=~0.30~\AA; the total of 0.7~\AA\
is slightly higher than that measured on December~2, probably due to
resolution effects.

\section{Analysis}

\subsection{Lightcurve Fits}
\label{sec:lcfits}

\begin{figure}
 \centering
 \includegraphics[width=0.760\textwidth,clip=true]{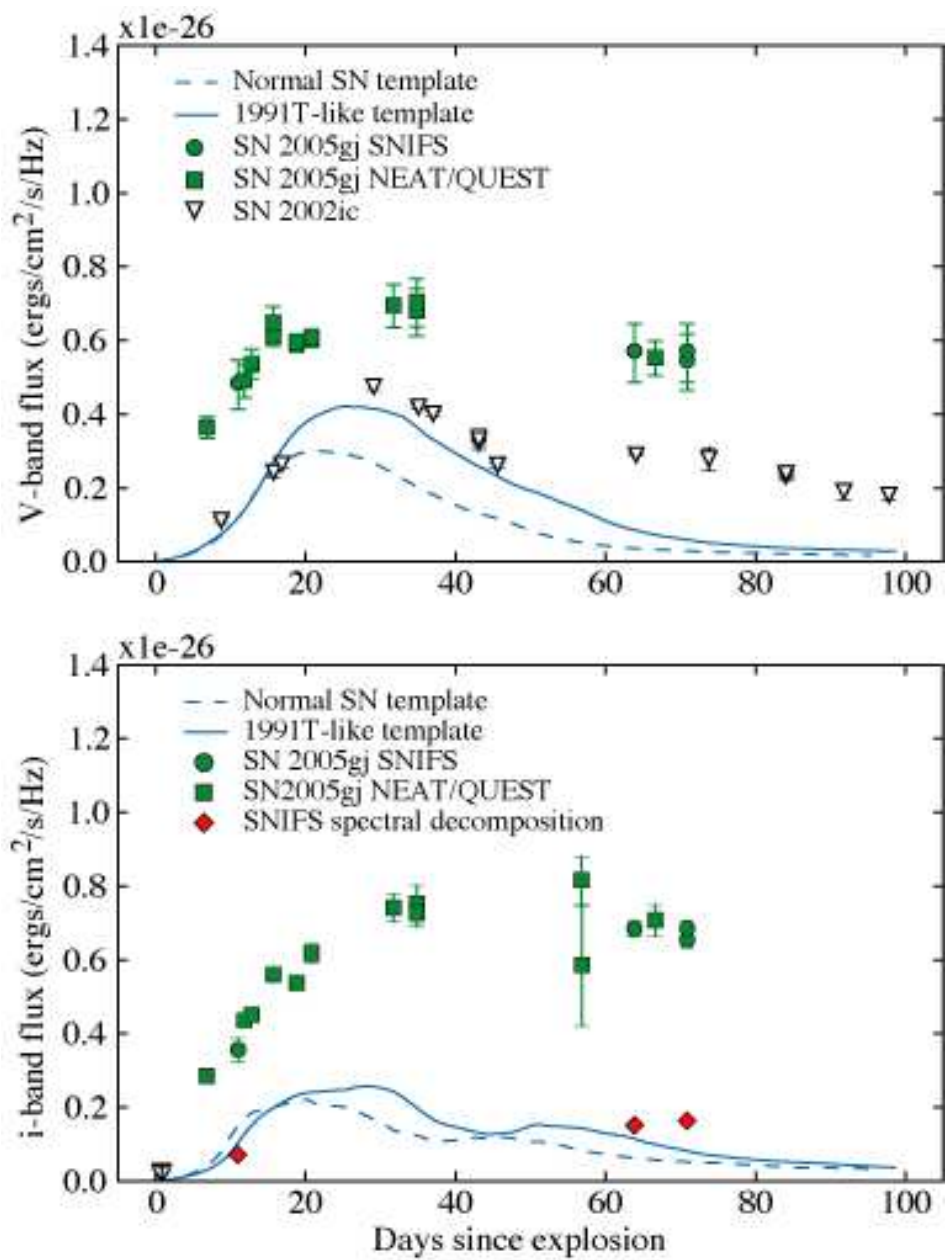}
 \caption{Comparison of \gj\ and \ic\ flux with \snia\ templates in
     \emph{V}-band (a) and SDSS \is-band (b).
     The \gj\ \emph{V}-band data are extrapolations
     from other filters based upon the shape of the SNIFS spectra.
     The inferred \snia\ flux from a spectral
     decomposition is shown with red diamonds.  To fit these
     points, the template needs to be brightened by 0.12 mag and either
     stretched by 1.32 or shifted by 16 days (or a combination thereof).
 }
 \label{fig:lightcurve_fit}
\end{figure}

\begin{figure}
 \centering
 \includegraphics[width=0.760\textwidth,clip=true,angle=90]{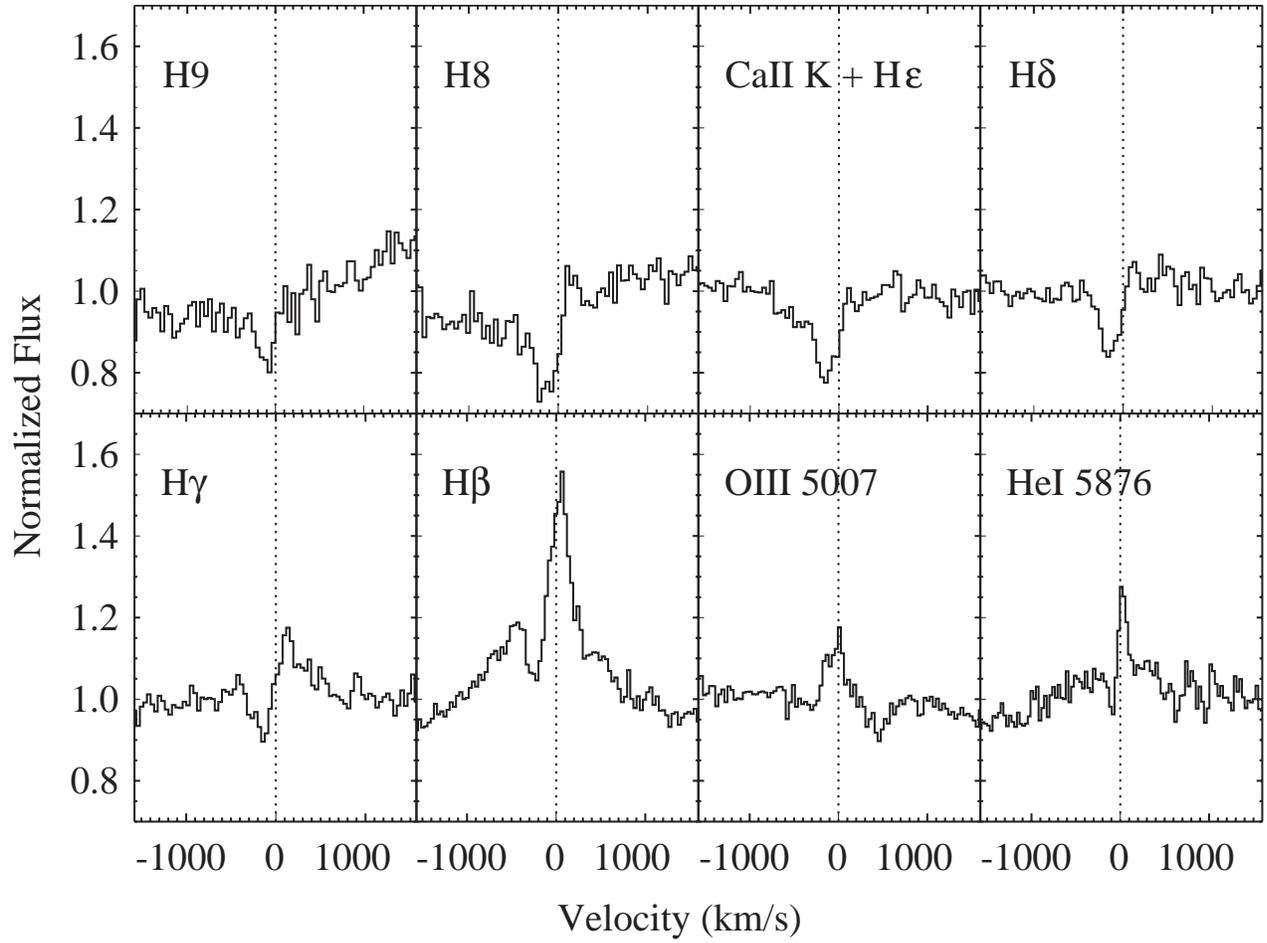}
 \caption{Velocity plot for hydrogen Balmer, [\ion{O}{3}]~\wl 5007, and
          \ion{He}{1}~\wl 5876 lines from the day 133 spectrum
          obtained with LRIS.
 }
 \label{fig:linezoom}
\end{figure}

Figure~\ref{fig:lightcurve_fit} shows a comparison between \gj,
\ic, and SN templates in Johnson \emph{V} and SDSS \is\ filters.
The SN templates used are new versions of templates discussed in
\citet{nug02}.\footnote{http://supernova.lbl.gov/$\sim$nugent/nugent\_templates.html}
The templates are scaled so that $M_B^{peak} = -19.5$ for the
Branch-normal case and $M_B^{peak} = -19.86$ in the \tlike\ template
\citep{sah01}.  In \emph{V}-band, the \gj\ points are synthesized
photometry from SNIFS observations, as well as values interpolated or
extrapolated using NEAT and QUEST data in other bands.  The \is-band
points include NEAT and QUEST photometry measurements as well as
synthesized photometry from SNIFS.  For comparison, we again show the
corresponding \emph{V} photometry for \ic.  In order to understand the
\gj\ lightcurve, we briefly review the modeling that has already been
performed for \ic.

Using only the \cite{ham03} \ic\ \emph{BVI} data for \ic, \cite{chu04b}
found that a flat CSM radial density profile was needed to fit the data
around 60 and 80 days after explosion.  \citet{uenishi04} found that
$\rho \propto r^{-1.8}$ was needed once the late-time \ic\ photometry
from \citet[][; not shown]{deng04} was included.  An updated model
by \citet{chugai04c} inferred the presence of a flat density profile
out to $3\times10^{15}$~cm, followed by a $\rho \propto r^{-1.4}$ fall
off. For this model it was noted that \snia-CSM interactions will have
a much more efficient conversion of kinetic energy to luminosity than
a \sniin\ due to the substantially higher X-ray opacity from Fe-group
elements in the \snia\ envelope \citep{chugai04c}.  \citet{woo04} added
several early-time photometry points, additional late-time photometry,
and photometry that better separated the SN lightcurve from a later
bump due to the CSM interaction.  They found reasonable agreement with
the data using the fainter \tlike\ template from \citet{nug02}, a CSM
density profile of $\rho \propto r^{-2}$, and a constant kinetic energy
to luminosity conversion term for the CSM interaction \citep{che94},
provided that the SN ejecta did not interact with the CSM until around
8~days after the explosion.  A more generalized --- but non-unique
--- density profile with excess material for the second bump was able to
fit all the data; however, an early gap was still required \citep{woo04}.

\begin{figure}
 \centering
 \includegraphics[width=0.760\textwidth,clip=true]{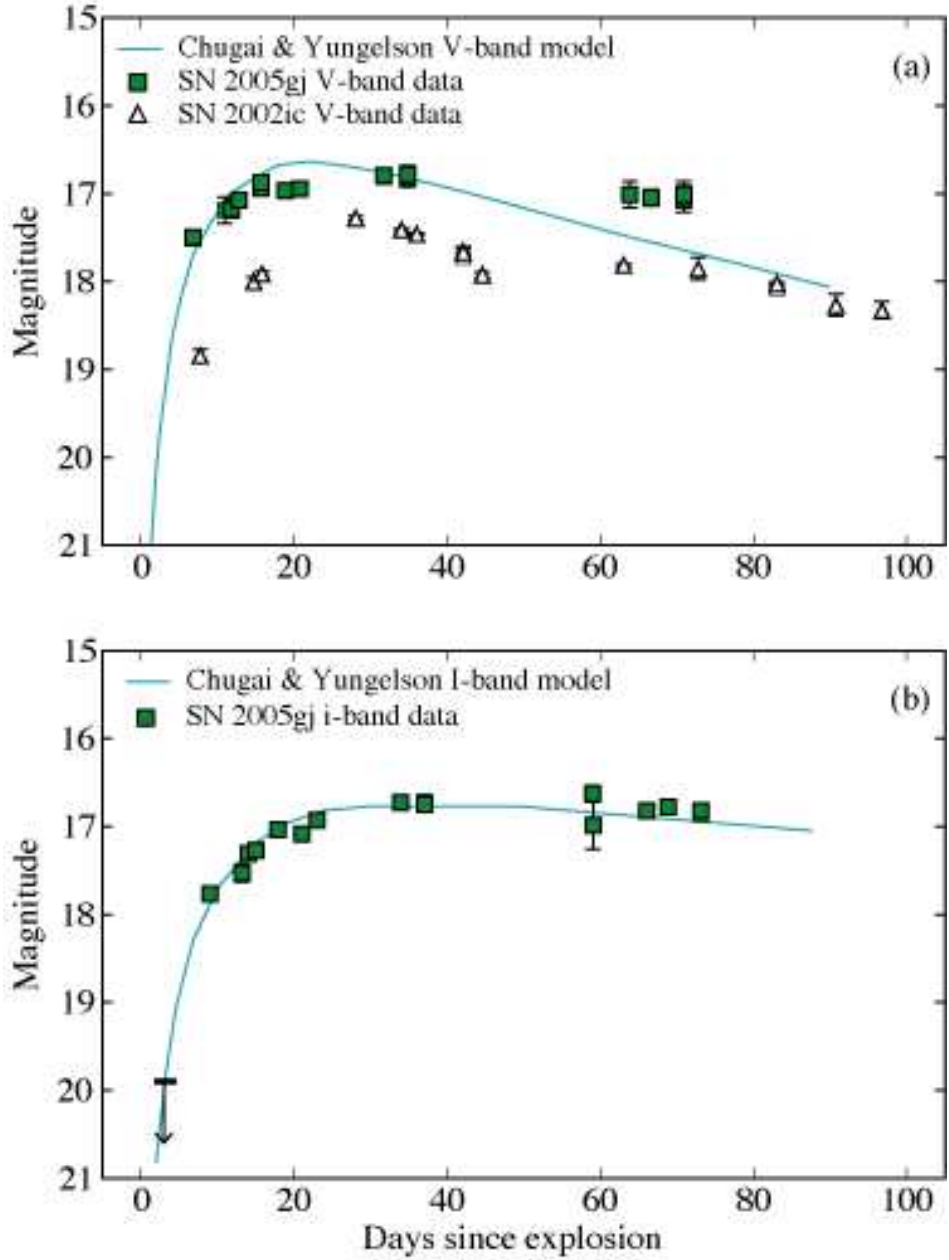}
 \caption{Comparison of \gj\ data with the \emph{V}- and \emph{I}-band
     lightcurve model of \cite{chu04b} as calculated for \ic\ using a flat CSM
     radial density profile.
     The \gj\ \emph{V}-band data are extrapolations
     from other filters based upon the shape of the SNIFS spectra.
     The magnitudes of the model have been shifted to align with the data
     and a best fit explosion date of 2005 September 20.4 is used.
 }
 \label{fig:chugai_lightcurve_fit}
\end{figure}

\gj\ is significantly brighter than either \ic\ or the SN templates,
indicating a much stronger and earlier CSM interaction.  Therefore,
unlike \ic, no decline after SN maximum light is seen. Over the time
period for which we have \gj\ data, the \cite{chu04b} model with a flat
density profile matches reasonably well in \emph{i}-band, as shown in
Figure~\ref{fig:chugai_lightcurve_fit}.  However, this model declines
too quickly in \emph{V}-band.  The best fit in \emph{i}-band is obtained
with an explosion date of 2005 September~20.4, which is 2.2 days prior
to the estimation from a simple second order polynomial fit to the rise
of the lightcurve.

Our first spectrum is from JD 2453647.09, which is 11 days after
explosion for an explosion date of September 22.6.
The expected fraction of SN light on that day is sensitive to the
assumed explosion date --- $29.1^{+14.3}_{-20.2}$\% in \is-band
and $24.6^{+11.5}_{-14.2}$\% in \emph{V}-band --- where the quoted
uncertainties are determined by adjusting the assumed explosion date
between September~18.6 and September~24.6.  On later dates for which
we have spectra the uncertainty due to the explosion date is much
smaller. For 64 and 71 days after explosion the \tlike\ lightcurve
template comparison implies \is-band \snia\ contributions of
$16.7^{+2.3}_{-1.7}$\% and
$12.2^{+2.4}_{-1.2}$\%, respectively.
In \emph{V}-band these percentages are
$14.9^{+4.7}_{-1.6}$\% and 
$14.9^{+2.1}_{-1.0}$\% .
The quoted uncertainties do not include
uncertainties in the lightcurve template itself.

If there were a gap between the \snia\ and the inner CSM radius
interaction, as found by \citet{woo04} for \ic, the \snia\ explosion
could have occurred as early as September 15.9 based upon the lightcurve
data alone.  However, the SN light fractions derived from spectral
decomposition in \S \ref{sec:spectral_analysis} imply a \snia\ explosion
date later than September~22.6. (As noted below, the radiation field
from the shock may modify the spectrum within the SN photosphere at
early times.)  For the purposes of this paper, we consider September~18.6
to 24.6 to be a conservative range of possible explosion dates.

\subsection{Spectral Analysis}
\label{sec:spectral_analysis}

Our first spectrum, shown at the top of Figure~\ref{fig:spectra}, was
observed 4 days after our discovery or 11 days after explosion, using
the outburst date of September~22.6 derived in \S\ref{sec:photometry}.
This is the earliest spectrum of any confirmed (\ic) or suspected
(SN~1997cy or SN~1999E) \snia$+$CSM event.  This initial spectrum of
\gj\ is indistinguishable from that of a classical \sniin.  The only
structure visible is the narrow \Halpha\ emission superimposed on a
broader component; there is no obvious indication of any feature typically
associated with a \snia\ spectrum.

\begin{figure}
 \centering
 \includegraphics[width=1.0\textwidth,clip=true]{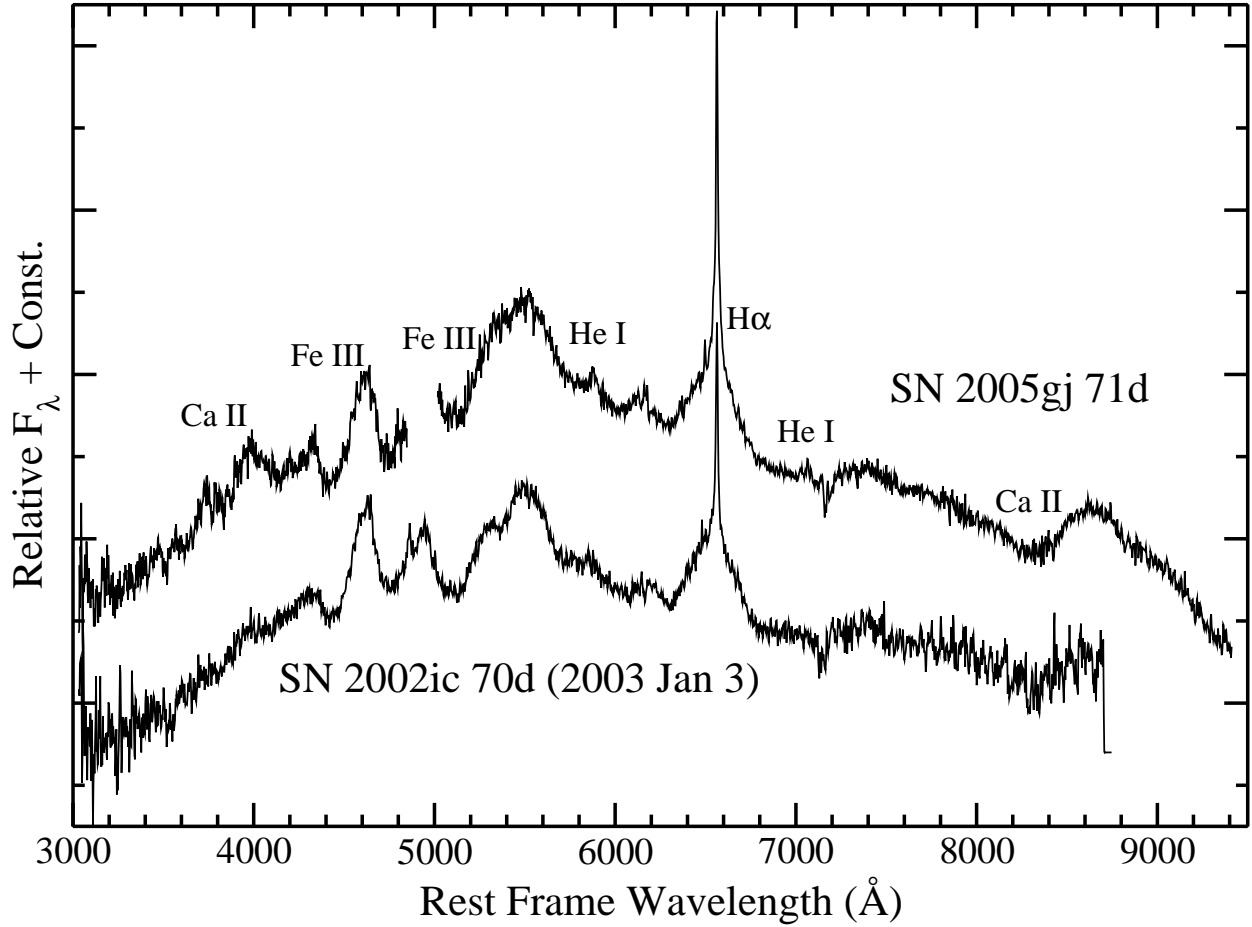}
 \caption{Comparison of the spectrum of \gj\ 71 days after explosion
 with that of \ic\ at approximately the same phase (assuming an explosion
 date of 2005 November~5~UTC for \ic).}
 \label{fig:late_comp_2002ic}
\end{figure}

Our later spectra, also in Figure~\ref{fig:spectra}, agree almost
feature-for-feature with the spectra of \ic\ at about the same phase.
Figure~\ref{fig:late_comp_2002ic} illustrates the overall similarity of
the combined SNIFS spectrum obtained 2005 December~2.5 (71~days after
explosion) with the 2003 January~3 spectrum of \ic.  A few differences
are evident in this comparison.  The spectrum of \gj\ possesses an
emission peak at 4000~\AA\ while that of \ic\ does not.  Also, the
absorption notch on the blue side of the emission hump at 5500~\AA\
clearly visible in \ic\ is much more subtle in \gj.

In their analysis of \ic, \citet{ham03} remark on the similarity of
their spectra to that of a \snia\ with ``diluted'' spectral features,
and decompose their spectra in terms of a \snia\ component plus a
smoothly-varying continuum.  This simple type of spectrum decomposition
in the case of SN$+$CSM interaction is not strictly correct, since it
neglects radiative coupling between the CSM interaction and line-forming
regions \citep{bra00,chugai04c}.  The CSM interaction region illuminates
the SN envelope from without and ``mutes'' the SN features.  P~Cygni
profiles lose contrast and may even invert, depending on details of the
CSM interaction such as luminosity and geometry.  A more detailed model
of the formation of the SN spectrum including the effects of this
coupling would help constrain the effects of muting, and also provide
insight into the geometry of the interaction region itself.  Still, a
simple linear decomposition of \gj\ may constrain the SN contribution in
the SDSS \is-band flux where \snia\ spectrum formation is less directly
influenced by line opacity \citep{pinto2000}.

\begin{figure}
 \centering
 \includegraphics[width=0.90\textwidth,clip=true]{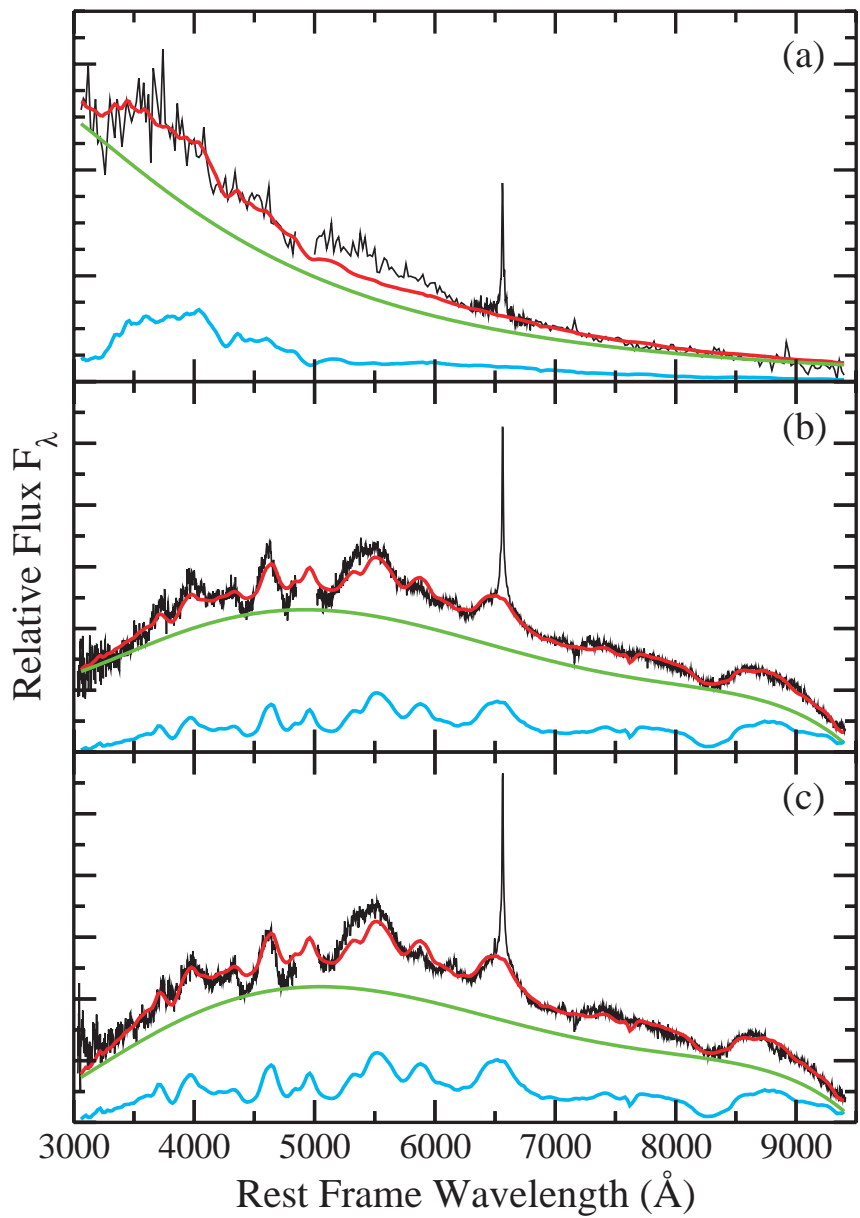}
 \caption{Decompositions of the \gj\ spectra.  Panel (a) compares the
 spectrum from 11 days after explosion (rebinned outside the \Halpha\
 region for clarity) to a 12,400 K blackbody plus \tlike\ template
 spectrum from the same rest frame phase.  Panels (b) and (c) show the
 decomposition of the day~64 and day~71 spectra in terms of a smooth
 polynomial plus \tlike\ template at 60 and 66 rest frame days after
 explosion (respectively).  In each panel the sum of the two components,
 and the two individual components, are shown.}
 \label{fig:decomp}
\end{figure}

After applying a Galactic extinction correction of \ebmv\ $ = 0.121$, we
decompose the SNIFS spectra of \gj\ using a \tlike\ spectral template
\citep{nug02} and a smoothly-varying function of wavelength to represent
the CSM.  The spectroscopically normal \snia\ template does not account
for the SN features as well as the \tlike\ template.  The CSM
contribution in the earliest spectrum is modeled as a blackbody, and in
the later spectra as a low-order polynomial.  Using a blackbody at late
time produces a less satisfactory fit than the polynomial.  The
decompositions appear in Figure~\ref{fig:decomp}.

In Figure~\ref{fig:decomp}a, we present the day~11 spectrum
decomposed into a 12,400~K blackbody plus \tlike\ spectrum 10 rest-frame
days after explosion.  The derived ratio of the \snia\ template to total
\is-band flux is 20\%.  Thus, despite the superficial resemblance to
a \sniin, \gj\ does show subtle but significant evidence for \snia\
spectral features.  In fact the amount of SN light inferred from
the linear decomposition is consistent with the prediction made in
\S\ref{sec:lcfits}.  This self-consistency suggests that at most
a small fraction of the SN photosphere can be obscured by the CSM.

Decompositions of the later-time SNIFS spectra appear in Figures~\ref{fig:decomp}b and
\ref{fig:decomp}c.  We have adopted template spectra at
rest frame phases of 60 and 66 days after explosion for the observations
on November~25.5 and December~2.5, respectively.  On both epochs, the
\tlike$+$CSM superposition exhibits a flux deficit at 5400~\AA\ relative
to the observed spectrum.  This absorption is the same robust notch
observed in the spectrum of \ic.  In their analysis of the spectrum of
SN~1991T, \citet{mazzali95} attribute this absorption to \ion{Fe}{2} \wl
5535 but do not fit it.  A lower iron abundance relative to \ic\ seems a
less likely explanation than muting effects; other iron-peak features in
\gj\ are just as strong as in \ic.  Aside from this feature, most of the
observed features are generally reproduced.  Roughly 22\% and 25\% of
the SDSS \is-band flux on these two days are accounted for by the
spectral template, irrespective of any adjustment to the template color.
This discrepancy with the light curve analysis indicates that the
template light curve may need to be brighter at this phase than the
simple scaling to $M_B^{peak} = -19.86$ allows.  

The implied \snia\ fluxes from the spectral decompositions are indicated
in Figure~\ref{fig:lightcurve_fit}.  To align the template with the
implied \snia\ fraction, the template would need to be 0.12 mag brighter,
and either stretched by a factor of 1.32 or shifted by 16 days (or a
combination thereof). 

Note that as an experiment we also fit the two later-time spectra allowing
the phase to float, and find that somewhat better decompositions result
from rest-frame phases around 53 days after explosion.  The CSM and
template are better able to reproduce the emission hump at 5500~\AA, but
the notch still persists.  This difference is more likely to reflect an
insufficiency of the spectral decomposition than a slower (or delayed)
evolution of the underlying \snia. We note that the SDSS \is-band flux
contribution is around 22\% at both epochs.

Now we consider the velocity structure, luminosities, and line strength
ratios of the H and He emission lines observed.  As a starting point,
each line was decomposed into a broad and narrow component using two
Gaussians.
The \tlike\ template, multiplied by a linear function, was included in
the fit to account for the broad (FWHM $\sim $13000~\kms) variations in
the background.  Where our spectra departs too strongly from the template,
as for the \Hbeta\ feature, the background was instead accounted for by
a second order polynomial.

\begin{figure}[htbp]
  \centering
  \includegraphics[height=0.6\textwidth, angle=90]{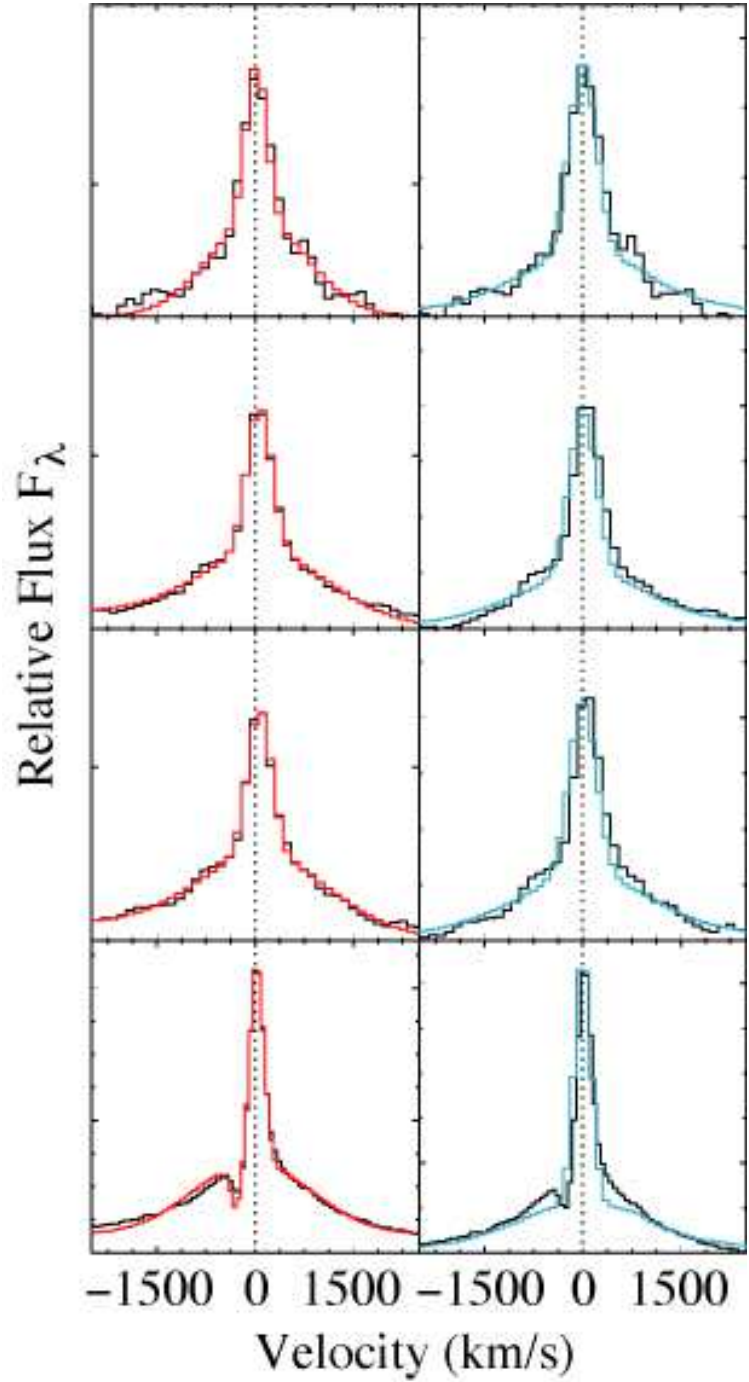}
  \caption{\gj\ \Halpha\ line velocity profiles and fits. Each row contains
  our measurements (black stepped line) on a given day. On the left, a model
  using two Gaussians (plus a P~Cygni profile for the last epoch)
  is shown (red line). On the right a fit based on a Monte-Carlo
  electron-scattering simulation is shown.
  The observations epochs are, from top to bottom,
  SNIFS day 11, SNIFS day 64, SNIFS day 71 and LRIS day 71.}
 \label{fig:Halpha}
\end{figure}

\begin{figure}[htbp]
  \centering
  \includegraphics[height=0.6\textwidth, angle=90]{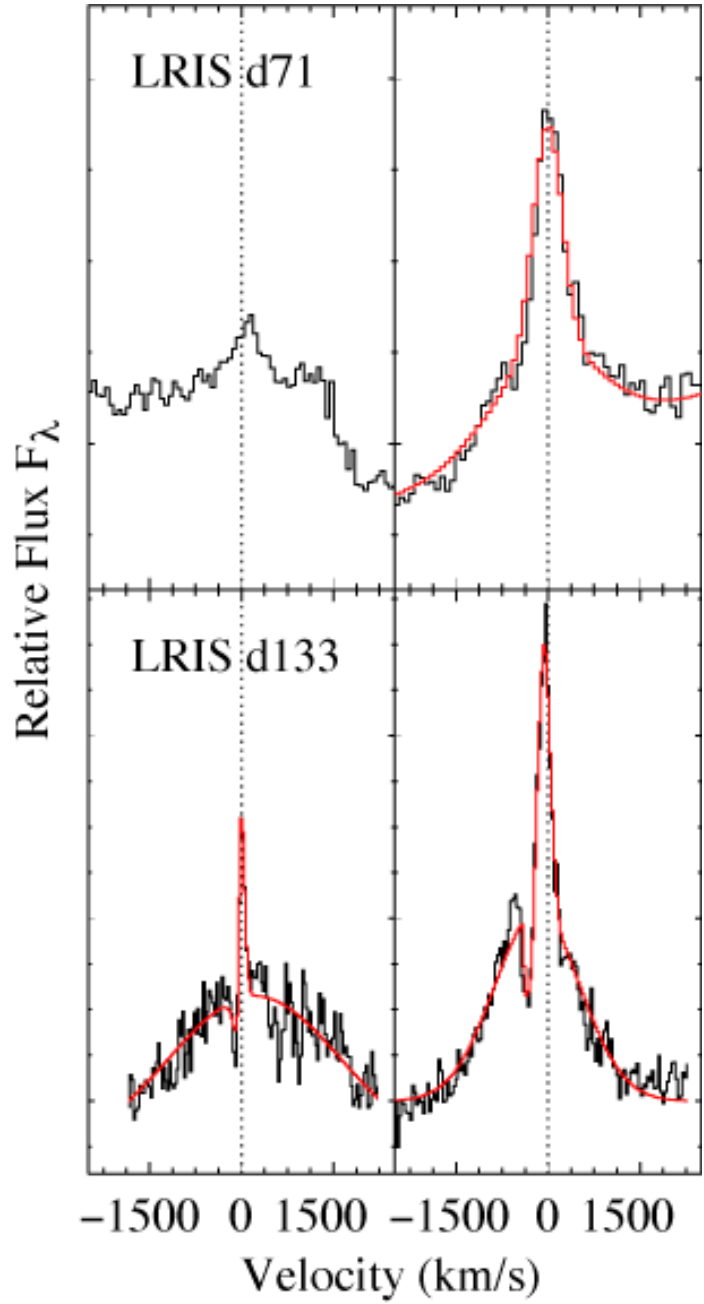}
  \caption{\ion{He}{1} 5876\AA\ (left) and \Hbeta\ (right) line velocity
  profiles and fits. First row shows LRIS day 71, with the \Hbeta\ line modeled
    as the sum of two Gaussians. Second row: LRIS day 133,  with both
    \ion{He}{1} 5876\AA\ and \Hbeta\ modeled as the sum of two
    Gaussians plus a P~Cygni profile.}
 \label{fig:HbetaHe}
\end{figure}

The results for the line widths and absolute luminosities
derived from these two-component Gaussian fits are tabulated in
Table~\ref{tab:TwoGaussFits} and the fits themselves are shown in
Figure~\ref{fig:Halpha} and \ref{fig:HbetaHe}.  The narrow and broad \Halpha\ features
remain constant in flux and width from day~11 to day~71. Both flux
variations are less than $\pm$ 20\% and their velocity variation is on
the of order $\pm$13\%. Further, the broad \Hbeta\ line width stays
constant from day~71 to day~133, suggesting that the widths of the
H lines do not evolve of a period of over 4~months. Note that
\Halpha\ from \gj\ is much stronger than that from the host.

As noted earlier, the \Halpha\ feature in the day~71 spectrum and the
\Hgamma, \Hbeta\ and \ion{He}{1}~\wl~5876 features in the day~133 spectrum
exhibit obvious P~Cygni profiles.  This morphology rules out an underlying
\ion{H}{2} region as an explanation for the narrow emission component.
We include a P~Cygni profile in our fits using a photosphere at $v_{phot}$
surrounded by a medium with \Halpha\ optical depth at the photosphere of
$\tau_{0}$, exponentially decreasing as $\exp((v-v_{phot})/v_{e})$. The
third free parameter is the density e-folding velocity ($v_{e}$).  The
results of these fits are given in Table~\ref{tab:PCygniFits} and the fits
are shown in Figures~\ref{fig:Halpha} and \ref{fig:HbetaHe}. Note that
there is strong covariance between the terms in the P~Cygni profile model,
and that the spectral resolution places a lower limit on the measurable
value of $v_{phot}$. Based on our high-resolution fit, we take $v_{wind}
\sim v_{phot} \sim$~\vw\ as characteristic of the radial velocity of the precursor wind.

Next, we analyze the inverted P~Cygni [\ion{O}{3}]~\wl 5007 profile seen
in the LRIS spectrum, shown in Figure~\ref{fig:linezoom}.  On day~71, the
profile has a net flux of zero, with a flux of $\sim 7 \times
10^{-17}$~erg~cm$^{-2}$~s$^{-1}$ and equivalent width of 0.28~\AA\
measured in both the emission and absorption components.  The blue and
red components are centered at $-100$~\kms\ and $+450$~\kms\ relative to rest
frame, respectively.\footnote{Due to slit effects and instrument
flexure, absolute velocities may be uncertain by $\pm 100$~\kms.
However, relative velocities are more certain, and are
therefore quoted to the nearest 50~\kms.}  By day~133, the emission has
increased in flux to $\sim 1.2 \times 10^{-16}$~erg~cm$^{-2}$~s$^{-1}$
while the integrated flux deficit in the absorption has decreased to
$\sim 5 \times 10^{-17}$~erg~cm$^{-2}$~s$^{-1}$ with an EW $\sim
0.34$~\AA.  The emission peak is shifted slightly, to $-50$~\kms, while
the location of the absorption remains unchanged.  The two components
have intrinsic widths much less than our 245~\kms\ spectral resolution.
This feature is discussed further in \S \ref{sec:csm}.

\section{Host Galaxy Properties}

The host galaxy of \gj\ is barely detectable in our
discovery images, however, it is detected in SDSS DR4
\citep{sdssdr4}, as shown in Figure~\ref{fig:sdsshost}, and in
the GALEX~DR1\footnote{http://galexgi.gsfc.nasa.gov/targets/DR1/}.
The galaxy is elongated, with an ellipticity of $\sim0.6$, and lacks a
strong central concentration.  The bulk of the galaxy (including the
location of \gj) is very blue (de-reddened $g$-$r = +0.18 \pm 0.10$,
$r$-$i = -0.02 \pm 0.13$) but the northern tip is extremely red according
to the SDSS \is\ images.  Weak \Halpha\ emission at the redshift of \gj\
is present at the location of this northern component in our Keck spectrum.
We have compiled the \gj\ host photometry in Table~\ref{tab:sdsshost}.
Using our measured heliocentric redshift of $z=0.0616$, the absolute
magnitude of the host is M$_B$=-17.4 or -16.9, depending on whether or
not the northern red component is included.  The luminosity-metallicity
relation for low-luminosity galaxies, recently updated by \citet{zsh06},
then suggests a host metallicity in the range $\rm{log(O/H)}+12=$~8.2---8.3.

The GALEX~NUV detection and the blue SDSS colors strongly suggest
that the light of a recent starburst dominates the blue host component
where \gj\ appears.  Thus, the age of the \gj\ progenitor system can 
most likely be
constrained by determining the age of the starburst. In order to constrain
the age of the starburst, we have fit the GALEX~NUV and SDSS $ugriz$
photometry to reddened starburst spectra from \citet{leitherer99}.
We chose the grid of instantaneous burst models for a metallicity of
$Z=0.008$ (supported by the low host luminosity and spectral analysis,
below). The \citet{cardelli89} extinction law was used to account for
Galactic extinction. In order to model the northern red component,
we selected a similarly red galaxy, seen in the lower left region of
Figure~~\ref{fig:sdsshost} and listed in Table~\ref{tab:sdsshost}, to form
a a red-component photometric template.  A joint fit for the amplitudes
of the red and age-dependent starburst templates yields a burst age of
$200 \pm 70$~Myr.  The fit has $\chi^2 = 1.05$ for 3 degrees of freedom,
indicating a match slightly better than the quoted photometry errors
would predict. The photometry, model components, and resulting fit are
shown in Figure~\ref{fig:Host_fits}.  This age estimate is sensitive
to the assumed amount of Galactic extinction, but can reach no higher
than 380~Myr for the case of no Galactic extinction.  To our knowledge,
this is the strongest constraint ever obtained on the age of a Type~Ia
supernova progenitor.

Our LRIS spectrum of the region of the host to the south of \gj\
shows strong \Halpha\, with EW(\Halpha)~=~35~\AA, indicative of modest
on-going star formation. Another striking feature of the host
spectrum is the absence of [\ion{N}{2}]~\wl 6584; this is an indicator
that the host galaxy metallicity is low \citep{denico02}.  However,
[\ion{S}{2}]~\wl 6717 and [\ion{S}{2}]~\wl 6730 are clearly visible.  In
order to set a better limit on [\ion{N}{2}]~\wl 6584, we simultaneously
fit the restframe 6500--6758 \AA\ spectral region, with Gaussian
profiles for the emission lines, a linear continuum, plus a scaled
version of the \gj\ spectrum in order to account for contamination
from wings of the PSF of the much brighter SN.  The emission lines
included in the modeling were \Halpha, [\ion{N}{2}]~\wl 6548, 
[\ion{N}{2}]~\wl 6584, [\ion{S}{2}]~\wl 6717, and [\ion{S}{2}]~\wl 6730, with
[\ion{N}{2}]~\wl 6548 set equal to 0.3 of [\ion{N}{2}]~\wl 6584.  A few
pixels badly affected by night sky lines were not included in the
fit, but these do not affect any of the regions containing host
emission lines. As the flux uncertainty in the remaining spectral
region is dominated by the featureless night sky, we assigned uniform
weights to each pixel. The weights were set to the inverse variance
of the sky, as measured directly from the spectrum.  As the wavelength
coverage of the fit is very small, no effort was made to account
for extinction by dust in this region of the host galaxy.

Our best fit matched the data within the assigned uncertainties.
Approximately equal amounts of the continuum were allocated to the
scaled spectrum of \gj\ and the linear continuum component. Thus, our
estimate of EW(\Halpha) given above suffers from pollution by \gj,
both in the \Halpha\ line and in the continuum. Our fit, which accounts
for contamination from \gj, gives  EW(\Halpha)~=~44~\AA. [\ion{N}{2}]~\wl
6584 is not detected; we are able to set a 95\% upper limit of
log([\ion{N}{2}]/\Halpha)\ $ < -1.24$. As [\ion{N}{2}]~\wl 6584 can also be
suppressed if the ionization parameter is high (see Figure~7 of
\citet{kewley02}, we also examined the \ion{N}{2}/\ion{S}{2} ratio,
which becomes an indicator of the ionization parameter at low
metalicities (see Figure~4 of \citet{kewley02}). We find a 95\% confidence upper
limit of log([\ion{N}{2}]/[\ion{S}{2}])\ $ < -0.61$; this indicates a very low
ionization parameter, or possibly an anomalously low abundance of N
relative to S.  N/S is found to have relatively modest scatter among
low-metallicity galaxies \citep{vanzee06}, so it is much more likely
that the ionization parameter is low in this case. For low values of the
ionization parameter, \citet{kewley02} show that \ion{N}{2}/\Halpha\ is
not very dependent on the ionization parameter.  Therefore, we can use
our upper limit on log([\ion{N}{2}]/\Halpha) to set a $2\sigma$ upper
limit on the metallicity of $\rm{log(O/H)} + 12 < 8.2$.  This is consistent with
the value expected based on the $B$-band luminosity
of the host galaxy, derived above, especially when the $\pm0.2$~dex scatter
in the luminosity-metallicity relation is included. Using the solar O abundance
determined by \citet{prieto01}, we find that the host of \gj\ has
$(Z/Z_{\odot})<0.3$. Deeper observations once \gj\ fades away will be
needed in order to determine a more accurate abundance.

Finally, it is worth examining the implications of
our Galactic and host \ion{Na}{1}~D measurements. The relation between
\ion{Na}{1}~D and \ebmv\ for stars in the Galaxy exhibits considerable
scatter \citep{herbig93,sembach93,munari97}. \citet{munari97} make
the distinction between systems showing \ion{Na}{1}~D at a single velocity
and those with multiple velocity components; this can affect whether
the line strengths arise from a single stronger component that may
saturate, or from the linear superposition of many weaker ---
potentially unsaturated --- lines. Within our resolution
we detect only single components for both Galactic and host 
extinction. Therefore, we may apply the fit given for \ion{Na}{1}~D$_1$ in
Table~2 of \citet{munari97} to estimate \ebmv\ $ \sim 0.2$ from the
Galaxy and \ebmv\ $ \sim 0.05$ from the host. An alternative prescription
based on a mixture of supernova types \citep{barbon90} is clearly
non-physical --- being linear up to column densities where \ion{Na}{1}~D
must be highly saturated. This may be due to issues with \ebmv\
derived from the colors of highly-extincted supernovae \citep{wang05a}.
Given the dispersion in the relation between \ion{Na}{1}~D and extinction,
we consider the agreement between the \citet{sfd} and \ion{Na}{1}~D
estimates for Galactic extinction to be reasonable. Based
on the weakness of the host (or CSM) \ion{Na}{1}~D, we
consider \gj\ to have negligible extinction from its host.

\section{Discussion}

A complete description of the geometry, hydrodynamics, and radiative
structure of the CSM surrounding \gj\ requires detailed modeling that is
beyond the scope of this paper.  However, our observations serve as
important constraints that any successful model of \gj\ must satisfy.
More generally, these observations can serve as tests for more generic
models designed to account for the apparent variation in the subclass of
CSM-interacting SNe.  

Here, we summarize our observations and then discuss them in terms of a
few crude, but standard, approximations.  The light curve of \gj\ rose
more quickly and to a higher peak luminosity than did \ic, and declined
much more slowly.  The earliest spectrum consists of a pseudo-blackbody
continuum punctuated by narrow \Halpha\ emission.  Spectra obtained two
months after outburst are similar to those of \ic\ at the same phase,
like those of a ``diluted'' \tlike\ event.  These later spectra also
include narrow and broad in \Halpha, \Hbeta,
\Hgamma, \ion{He}{1}~\wl5876, and \ion{He}{1}~\wl7065, with all but
\ion{He}{1}~\wl7065 displaying P~Cygni profiles.
Blue-shifted higher-level Balmer lines in absorption, each possessing a ``shelf''
feature to the blue, are observed.  The widths of the broad \Halpha\ and
\Hbeta\ components do not change with time, the narrow \Halpha\
luminosity is also fixed, and the observed Balmer decrement is large.
Further, we observe a --- to our knowledge unique ---
inverted P~Cygni [\ion{O}{3}]~\wl5007 line.

\subsection{CSM Density, Mass, Size and Geometry}
\label{sec:csm}

The early brightness of the light curve relative to that of a \snia, the
blackbody nature of the spectrum on day~11, and the agreement with the
\citet{chu04b} model light curve indicate that \gj\ began to interact
with its CSM within 7~days of explosion.  This time scale and a
reasonable estimate for the outer SN ejecta velocity set an upper limit
on the inner radius of the CSM.  Ejecta velocity at the photosphere of a
typical \snia\ 7~days after explosion is 23,000~\kms, later dropping to
6,000~\kms\ \citep[see Figures~10 and 11 of][and references
therein]{garavini05}, so we adopt an outer ejecta velocity of $v_{SN} =
25,000$~\kms\ and infer a maximum inner CSM radius of $R_i^{max} < 1.5
\times 10^{15}\ ( v_{SN} / 25,000~\kms)$~cm.

While our earliest spectrum allows \snia\ radiation to be present, the
overall spectral shape is reasonably accounted for by a blackbody.  By
scaling an undiluted $T = 13,000$~K blackbody to our day~11 spectrum we
arrive at a bolometric luminosity of \lbol\ at this phase, corresponding
to a radius of $R_{BB} \sim 2.9 \times 10^{15}$~cm.  The velocity needed
for the SN ejecta to reach this radius in 11 days is $\sim 30,000$~\kms.
If the explosion date is moved 2 days earlier --- in agreement with the
\citet{chu04b} light curve fit in the \is-band --- the required ejecta
velocity decreases to 25,000~\kms, in agreement with expectations.  This
appears to confirm the basic picture that the continuum emission arises
from the ejecta/CSM interface.  Of course, a spherically symmetric
(optically thick) blackbody interaction region is at odds with the
detection of the \snia\ component at day~11.  It seems that an opening
in the interaction region is required for the \snia\ to be seen.

The persistence of strong emission from \ion{He}{1} and hydrogen Balmer
lines through day~133 may be used to constrain the minimum outer CSM
radius.  The presence of narrow \Halpha\ emission and lack of broad
\Halpha\ absorption indicate that the broad hydrogen emission is powered
by the shock rather than radioactive decay \citep[][Table~3]{chugai91}.
It has become standard to identify the broad emission component with a
cooling shock front at the interface between the SN ejecta and CSM.  We
associate the broad features with the forward shock moving through the
hydrogen-rich CSM rather than with the reverse shock propagating into
the hydrogen-deficient SN ejecta.  Based on the measured velocity widths
of the broad component in Table~\ref{tab:TwoGaussFits}, we take $v_s =
$~\vs\ as representative of the shock velocity (but see discussion
below).  This is substantially lower than the shock velocity of
11,000~\kms\ derived by \citet{wang04} for \ic\ one year after
explosion, but in line with the 2,900~\kms\ inferred by \citet{kotak04}.
Hence, we estimate a minimum outer radius of $R_o^{min} > 2.8 \times
10^{15}\ ( v_s / \vs)$~cm.

Note also that the width of the narrow component is much greater than
any plausible thermal velocity.  These considerations and the presence
of a P~Cygni profile suggest that the narrow component is associated
with unshocked CSM photoionized by the shock.  In this case, the
broadening could be due to radial expansion of the unshocked material.
As the emission may arise from a precursor wind, we denote this velocity
as $v_w$, and use $v_w = $~\vw\ based on the velocity of the
P~Cygni profile.  It is possible that the observed value of $v_w$ is
higher than the historical value of a precursor wind if the pre-shock
region is accelerated due to photoionization heating \citep{chev97}.
For a constant $v_w$, the derived inner and outer CSM radii imply an
outflow beginning at least 15~years ago, and ending less than 8~years
prior to explosion.

The observed line strengths may also offer clues to the properties of
the CSM.  \gj\ exhibits a large Balmer decrement --- \Halpha/\Hbeta\
$\sim 6$ on day~71.  If this were due to dust, as is normally assumed
when the Balmer decrement departs from case~B recombination, \ebmv\ $\sim
0.7$ would be required.  This would be at odds with our failure to detect
strong \ion{Na}{1}~D absorption due to the host galaxy.  This much dust
would require 7 times greater optical luminosity, and in view of the
amount of SN light required by our spectral decompositions, the \snia\
required would be unphysical.  Moreover, for so much dust there would
be very poor agreement between the blackbody radius and the SN ejecta
radius at day~11.

Instead, it is more likely that the H level populations are not in
case~B. The Balmer decrement can be greatly enhanced at high density
when the optical depth in the \Halpha\ line is high. \citet{drake80}
demonstrate this effect for a static slab of H, and attribute it to a
combination of Balmer self-absorption (i.e., in which resonant trapping
leads to \Hbeta~$\rightarrow$~\Halpha~$+$~P$\alpha$) and collisional
excitation \citep[also see][]{xu92}.  The SN-CSM interaction models
of \citet{che94} show this effect, giving \Halpha/\Hbeta\ ~$\sim7$
after 1~year.  Further support for a high-density region comes from
the presence of strong \ion{He}{1}~\wl 7065 and the lack of evidence
for \ion{He}{1}~\wl 4471 or \wl 6678 at day~71.  The calculations by
\citet{almog89} suggest that a minimum electron density of $n_e \sim
10^{10}$~cm$^{-3}$\ is required to produce such \ion{He}{1} line ratios.
Clearly, the \gj\ line ratios offer a number of interesting clues
concerning the nature of the CSM.

As noted in \S\ref{sec:spectral_analysis}, a double Gaussian plus
P~Cygni profile provides an excellent fit to the \Halpha\ and \Hbeta\
lines.  The fits to \ion{He}{1}~\wl 5876 and \ion{He}{1}~\wl 7065
prefer a flatter, red-shifted broad component.  The question arises
as to whether or not these line profiles are consistent with a broad
component from a shock front and a narrow component in a photoionized
outer region. \citet{fransson84} has calculated expected line profiles for
emission in SN--CSM shocks. The shapes can be quite complex depending on
whether pure scattering dominates over collisional effects, but generally
have a roughly parabolic shape or a relatively flat-topped shape. Another
case of interest would be that of a shock propagating through a disc,
which should give a two-horned profile.  None of the example situations
give the symmetric Gaussians that we see for the Balmer lines.  Indeed,
one of the puzzles here is why the broad component would be symmetric at
all given that the redshifted portion of the CSM interaction should be
occulted by the \snia\ or the optically-thick foreground CSM.  On the
other hand, the broad components of the \ion{He}{1} lines are weak,
and probably asymmetric and flat-topped.  Therefore, it is unclear how
much Balmer emission is co-spatial with the \ion{He}{1} lines.

However, Thomson scattering of the lines also must be considered
\citep{chugai01,wang04}.  A simple Monte Carlo simulation shows that
for modest optical depth Thomson scattering produces a line profile
consisting of a narrow core from unscattered photons and broad wings
due to the scattered photons.  The width of the broad component and the
ratio of the broad and narrow Thomson-scattering components depend on the
electron velocity, $v_e$, and the scattering optical depth, $\tau{_e}$.
As most of the difference between a Thomson-scattering line profile
and a simple double-Gaussian occurs in the wings, the undulating SN
component makes it difficult to discriminate between these two cases.
For a $13,000$~K blackbody, $v_{e}\sim760$~\kms, in rough agreement
with the width of the broad component of \Halpha.  Our fit to \Halpha\
using Monte Carlo calculations gives $\tau{_e}\sim1.9$.  An electron
density of $n_e\sim10^{10}$~cm$^{-3}$ and a length scale of order
$l\sim2\times10^{14}~\rm{cm}$ gives an Thomson scattering optical depth
that agrees with these estimates.  This shows that Thomson scattering
alone could account for the observed broad hydrogen features for
parameters consistent with a shock.

A Thomson-scattering origin for the line widths suggests that both the
broad and narrow components of the H emission lines could arise from a
single, dense region.  This could help to explain why the Balmer decrement
is large for the narrow component as well as for the broad component;
\Halpha/\Hbeta\ for the narrow components would be expected to have
close to the case~B ratio if they arise, e.g., from photoionization of a
protoplanetary nebula. A single origin would also explain why the ratio
of the narrow and broad components is not changing appreciably with time.
Here the FWHM of the emission lines would no longer be related to $v_s$.
In this picture there is still material traveling at $v_w$ in front
of the shock, since blue-shifted, P~Cygni-like absorption is seen,
but a low-density photoionized region from a precursor wind would not
be required.

Emission from \Halpha\ will be heavily weighted towards high density
regions.  Assuming regions of roughly constant density, possibly having
a filling factor less than unity, we may write the \Halpha\ luminosity
as: 
\begin{equation}
  \label{eq:LifanHLum} 
  L_{\Halpha} = h \nu_{\Halpha} \alpha_{eff} n_{e} M / ( 1.4  m_{H} )
\end{equation} 
where $M$ is the mass in the shock region and $\alpha_{eff}$ is the
effective H recombination coefficient.  Here we assume that the emitting
region is 90\%~H and 10\%~He by number, and that He$^+$ and He$^{++}$
are both present.  \citet{wang04} took $\alpha_{eff}=\alpha_{B}$, the
case~B recombination coefficient. However, \citet{chugai90} advocates
$\alpha_{eff}=\alpha_{C}$, the case~C recombination coefficient, when
the Balmer optical depth is large, as is the case for the \gj\ CSM.
Taking $n_e\sim10^{10}~\rm{cm}^{-3}$ we find that $M\sim 1.6\times
10^{-2}\ (10^{10} \rm{cm}^{-3}/n_e)\ M_\odot$\ is required to explain the
observed \Halpha\ luminosity. Note that this mass estimate applies only to
the mass in the emitting region.  

The measured $L_{\Halpha}$ and inferred Thomson optical depth can be
used to set more direct constraints on the dimensions of the emitting
region.  For example, for $n_e\sim10^{10}~\rm{cm}^{-3}$ and a radius
$R\sim 2\times 10^{15}$~cm, $1.6\times 10^{-2}\ M_\odot$ would be
contained in a spherical shell with $\Delta R \sim 3\times10^{13}$~cm.
Such a shell would not be optically thick due to electron scattering
and is therefore inconsistent with our measurements.  If instead
$n_e\sim10^{9.0}~\rm{cm}^{-3}$, then the \Halpha-emitting gas would
be $0.16\ M_\odot$ and the shell radius would be $\Delta R \sim
3\times10^{15}$~cm and the emitting region would be optically thick
to Thomson scattering. Alternatively, as the visibility of the SN
requires the presence of an opening in the shock front, we can imagine
the emitting region as a working its way through a circumstellar disk.
If the emitting region were at a radius $R\sim 2\times 10^{15}$~cm with
mass $0.016\ M_\odot$ the emitting region would have characteristics
dimensions of $\sqrt{l \Delta R} \sim 3\times10^{14}$~cm, where $l$ is
the thickness. This region would be optically thick to Thomson scattering
and this configuration would allow the \snia\ to be easily seen.

More generally, we may combine these basic ingredients to give the value
of $n_e$ that will produce the correct Thomson optical depth and \Halpha\
luminosity from a region with characteristic size $l$:
\begin{equation}
  \label{eq:GregNe}
  \bar{n}_e = 0.9\ h \nu_{\Halpha} \alpha_{eff} (\tau_e/\sigma_T)^3 L_{\Halpha}^{-1}
\end{equation}
We find $\bar{n}_e\sim1.5\times10^{8}$~cm$^{-3}$ and
$l\sim2\times10^{16}$~cm. The total mass in the emitting region
would then be $\sim0.9\ M_\odot$ This analysis disagrees with our
density estimate from the H and He line ratios and the sizes inferred
from our SN ejecta speed or blackbody luminosity calculations. 

However,
if the emitting mass is in the form of $N_{cld}$ equal clouds, the value
of $\bar{n}_e$ increases in proportion to $N_{cld}$, $l$ decreases
proportionately in order to maintain the electron scattering optical depth
for each cloud, and the total mass over all clouds decreases as $N_{cld}$.
To reach $n_e \sim 10^{10}$~cm$^{-3}$ requires $\sim 100$ clouds.
Such clouds could be similar to fingers from Rayleigh-Taylor instability
that produce a corrugated SN-CSM interface in the \citet{chugai04c} model 
for \ic. Motion of the clouds could contribute to the Gaussian shape
of Balmer line profiles, possibly decreasing the electron-scattering
optical depth we have inferred. 

While electron scattering is able to explain the line profiles we
observe, it is worthwhile considering the possibility that the electron
scattering optical depth is low. In this case the broad component would
be due to the kinematics of the shock.  In the thin-shell approximation
\citep{chev82}, $v_s$ evolves as $t^{(s-3)/(n-s)}$, where $n$ is the
power-law index of the density profile of the outer SN ejecta and $s$
is the power-law index of the CSM material. While \snia\ outer density
profiles are not thought to follow a power law, $n\sim7$ is often taken
as an adequate approximation for application of the self-similar solution
of the thin-shell approximation. The constant CSM density, $s\sim0$,
profile inferred from the lightcurve then gives $v_s \propto t^{-3/7}$. We
have measured the Balmer line widths spanning a factor of 10 in time
and do not detect any significant change in those widths. The only ways
to maintain a constant $v_s$ are to have $s\sim3$ -- at odds with the
lightcurve -- or to have an unphysically steep, $n\rightarrow\infty$,
outer density profile for the SN ejecta.  Thus there is an apparent
inconsistency with the classic shock model explanation for the
broad component

Further aggravating this inconsistency is the constancy of the Balmer
emission.  In the shocked region itself, the \Halpha\ luminosity is
proportional to the time rate of dissipation of kinetic energy at the
shock front, $$L^{\textrm{\tiny{Broad}}}_{\Halpha} \sim 1/2 \dot{M}_{CSM}\
v_s^2$$ where $\dot{M}_{CSM}$ is the rate at which the CSM is swept
up by the shock \citep{chugai91,sal98}. Therefore the stability of
$L^{\textrm{\tiny{Broad}}}_{\Halpha}$ requires a very fine tuning between
$\dot{M}_{CSM}$ and $v_s$. This is true for any shock, whether or not
the broad component is due to electron scattering.  In our discussion
of electron scattering we considered the case of a large population
of clouds. Here, the shock would interact with different clouds at
different times, substantially smoothing the emission-weighted $v_s$
and $L_{\Halpha}$.

While determining the exact shape, hydrodynamic state and radiative
structure of the CSM would require detailed modeling beyond the scope of
this paper, our observations combined with some rough approximations
have allowed us to infer that the CSM has a high density and that the
optically thick emitting region cannot significantly cover the \snia.
A high density is also consistent with the absence of detectable X-ray
emission \citep{imm05}, although the large X-ray opacity of the \snia\
envelope could be equally important \citep{chugai04c}.  Likewise, the
null detection in the radio \citep{sod05} may be due to absorption by
thermal electrons in the dense CSM.  Line absorption from a circumstellar wind
is detected extending out to our last date of observation.  The good
agreement of the lightcurve with a flat density profile suggests that
the current interaction is not with a classical $\rho \sim r^{-2}$
wind. Although there is no evidence for delayed CSM interaction, as was
seen for \ic, it is possible that a fast wind from the white dwarf has
piled up material where the interaction is now occurring \citep{chev97}. Alternatively,
the mass loss may have tapered off with time, producing a flatter density
profile. The mass of the emitting region is quite small if the densities
are as high as we have inferred. Unlike the classical case where the
narrow Balmer emission arises from the photoionized precursor wind,
in our picture we can not use Equation~\ref{eq:LifanHLum} to place
a lower limit on the mass in the wind. Therefore, the total CSM mass
remains uncertain, and will have to be estimated using a self-consistent
radiative hydrodynamic model.

One remaining clue to the nature of the SN-CSM interaction is the
inverted P~Cygni profile of [\ion{O}{3}]~\wl 5007 seen in the Keck
spectra on days~71 and 133.  The inverted P~Cygni profile morphology
is extremely interesting, and could indicate that the line forms in
a region that is excited from the outside rather than from the SN,
and forms in the intershock region.  A very high density is required to
achieve sufficient optical depth for absorption in the line.  The oxygen
abundance of high-velocity \snia\ material should be much higher than
that found in the CSM \citep{hatano99} and therefore we conclude that the
feature is most likely to originate within the SN ejecta.

\subsection{Progenitor Models}

One of the potentially powerful applications of rare events like \gj\ is
in understanding the progenitor systems of \sneia. While the interaction
with the CSM provides many clues to the progenitor system, the host
galaxy environment adds additional information and helps to place such in
events in the context of star formation and evolution models.  \gj\ is a
rare example in which observations of the host galaxy place interesting
constraints on the likely progenitor age and metallicity.  In this section
we explore the broader implications of our host-galaxy measurements.

If the progenitor system of \gj\ shares the approximately $200$~Myr age
of the starburst dominating the host galaxy, this places some constraints
on the allowed parameters of the progenitor system.  A 3~M$_{\odot}$
star -- thought to be the minimum mass for forming a Type~Ia from a
carbon-oxygen white dwarf -- will take 400~Myr to complete its He-burning phase
according to the evolutionary models of \citet{schaerer93} for stars
with $Z/Z_{\odot} = 0.4$. A 4~M$_{\odot}$ star requires only 190~Myr. Thus,
the inferred age of the starburst is quite consistent with the mass range
of \snia\ progenitors derived from other stellar evolution calculations
\citep[see][and references therein]{umeda99}.
Simulations by \citet{belczynski05}
indicate that, for their fiducial model parameter settings, both
single-degenerate and double-degenerate channels will just barely
have become operative 200~Myr after birth of the progenitor system.
Better agreement with a 200~Myr timescale is obtained by decreasing
the CE efficiency factor.  Broadly speaking, single-degenerate
progenitors tend to have timescales that are strongly clumped in the 200
to 800~Myr timescale, while double-degenerate progenitors tend to occur
more uniformly over a 100~Myr to 10~Gyr timescale. Thus, the age of the
host starburst is consistent with modeled timescale estimates for both
single-degenerate and double-degenerate progenitors.  On the other hand,
if \gj\ is associated with the old stellar population in its host galaxy,
the \citet{belczynski05} models suggest that the double-generate scenario
would dominate after several Gyr.

The low metallicity derived for the host galaxy may be important
in understanding \gj.  The metallicity is likely to have an impact
on how quickly the winds from the progenitor or donor star can
be swept out of the system via radiation pressure from a hot WD,
with low metallicity increasing the likelihood that CSM will be present when the \snia\
eventually explodes.  Extreme metallicities may even lead to new
channels for thermonuclear SNe. For example, \cite{lowmetalSN} argue
that metal-free ([Fe/H]$<-5$) stars in the range $3.5M_{\odot} < M <
7M_{\odot}$ could explode as thermonuclear SNe inside a dense unburned
hydrogen envelope. This is close to the Type 1.5 scenario where the
carbon-oxygen core of a single massive asymptotic giant branch star
undergoes a thermonuclear explosion \citep{ham03}.  While we have only
an upper limit on the likely metallicity of the progenitor system based
on the host galaxy, it seems unlikely that the metallicity is as low
as required in this scenario.  \citet{zijlstra04} has suggested that a
low mass loss rate from an AGB star due to low metallicity could allow
the carbon-oxygen core to reach the Chandrasekhar limit, leading to a Type~1.5.
An example is a star with $\textrm{[Fe/H]}=-1$ and an initial mass of
6.5~$M_\odot$; such a star would complete its evolution in $\sim 60$~Myr
\citep{schaller92,schaerer93}.  A progenitor star of such mass is unlikely given
our starburst age estimate, but a progenitor of slightly lower mass
would be allowed. In the \citet{zijlstra04} scenario, this would require
$\textrm{[Fe/H]}<-1$, and would require that the metallicity of the host
of \gj\ fall significantly below the standard galaxy luminosity-metallicity
relation.

More apropos would be the studies such as \citet{kobayashi98, kato99}
examining the relation between progenitor metallicity and accretion onto
a classical carbon-oxygen WD.  They find that at low metallicity the deposition
rate is too high due to the reduced radiation pressure from the WD on
the infalling material; low metallicity systems thus fail to produce Type~Ia
events. \citet{kobayashi98} present the allowed values of donor mass
and orbital period for solar and 0.2$\times$ solar abundance for a main
sequence donor and red giant donor. Our age limit from the host starburst
precludes their 1~$M_\odot$ red giant as a donor, while our upper limit of $Z/Z_\odot < 0.3$
shrinks the allowed mass of the main sequence donor in their models to
$\sim 2.2\pm0.4\ M_\odot$ (and an orbital period in the range $0.4<P<4$ days;
the range of separation between the donor and a white dwarf of
1.4~$M_\odot$ placed by these limits is $2.3\times10^{11}$cm $<a<
1.7\times10^{12}$cm). 

Other possibilities include the \citet{livio03} model for \ic, in which
the merger of two white dwarfs immediately follows a common envelope
(CE) phase. In this model, the merger due to gravitational radiation
occurs after a common envelope stage has finished expelling mass from
the system.  \citet[][eq.~(3)]{chu04b} argue against this scenario in the
case of \ic\ because the white dwarf and core cannot get close enough
to merge via gravitational radiation on a timescale commensurate with
the presence of CSM. Getting the white dwarf and core sufficiently close
would require removal of energy via the envelope, but the observed CSM
velocities are simply too low. This is also the case for \gj. We infer
that mass loss ended at most 8 years prior to explosion, so as with \ic,
the timescale for gravitational wave radiation after the CE phase is
very short. Likewise, the velocity required for removal of sufficient
energy via the envelope is too low by well over an order of magnitude.

Perhaps more interesting in the case of \gj\ is the scenario recently
described by \citet{han06} in which a delayed dynamical instability
operating over a period of $\sim10^4$~yrs allows a $\sim3\ M_\odot$
donor star to eject $\sim2\ M_\odot$ of material out of the binary
system while a trickle of mass builds up an already massive WD to the
point of thermonuclear explosion. While constructed to explain \ic,
this model seems to be consistent with what we have been able to infer
from our observations of \gj. More detailed modeling of \gj\ will be
needed, however, in order to test the \citet{han06} scenario.  This model
gives in a predicted incidence of CSM-interacting \sneia\ in the range
0.001---0.01, which agrees within the large uncertainties with the 2-4
such systems discovered over the past 10 years in which $\sim550$ nearby
\sneia\ have been reported.  A prediction of the \citet{han06} model ---
which did not arise directly from \ic\ --- is that the highest incidence
of such events will occur 0.1--1~Gyr after the birth of the progenitor
system; our age estimate for \gj\ falls squarely within this window.

In a cosmological context it is interesting to ask whether a \snia\
like \ic\ or \gj\ could mistakenly be included among a sample of normal
\snia\ used to fit for the cosmological parameters. At high redshift the
\Halpha\ emission feature in our day~11 spectrum would be redshifted
out of the typical observing window. Thus, an observer would see the
smooth spectrum of a hot blackbody, likely superimposed on the spectrum
of a blue host galaxy, and probably classify the event has an early,
overluminous \snii. Several examples matching this description are shown
in \citet{lidman05}. At later times, when the classical \snia\ features
are apparent, the diluting affects of the CSM emission could be mistaken
for host galaxy light, e.g., from a starburst, resulting in a \snia\
classification \citep[See][for descriptions of the standard spectral
classification techniques applied at high redshift]{lidman05,hook05}.
A much stronger discriminant is the lightcurve shape. In the case of \gj,
the flat lightcurve would be a clear indicator of a serious problem. However,
as can be seen in \citet{woo04} for the case of \ic\ --- for which
the CSM interaction is weaker and its onset delayed --- lightcurve
data at least 40 days after maximum light would be needed to detect a
problem. Such data usually was not available in the Supernova Cosmology
Project or High-Z Supernova Search Team \snia\ cosmology searches of the
late 1990's.  More modern rolling searches such as the Supernova Legacy
Survey \citep{astier06} and ESSENCE \citep{kris05} contain such data
except for \sneia\ discovered late relative to the observing window for
a given program field. As the high-redshift \snia\ searches have netted
several hundred events to date, events like \gj\ and \ic\ should have been
detected unless their incidence declines strongly with redshift. In fact,
the opposite trend is expected, as the type of low-metallicity starburst
likely to have spawned \gj\ should be more prevalent at high redshift.

\gj, \ic, and SN 1997cy all exhibit large
CSM interactions and arose in low-luminosity hosts.  This underscores
the importance of blind searches that are not biased towards bright
known hosts for surveys that aim to provide data for understanding
\snia\ physics.  It also raises questions about the consistency between
the currently available nearby SNe sample (taken mostly from targeted
galaxy searches) and the distant SNe sample (taken primarily from untargeted
searches).

\section{Conclusions}

First classified as Type IIn and then displaying clear Type Ia features,
\gj\ is the second supernova spectroscopically confirmed to possess Type
Ia features combined with a strong CSM interaction. Of these two confirmed
events and two suspected such events, \gj\ presented the largest observed
CSM interaction, and its early behavior raises the question of how many
of the early detected \sneiin\ might be similar events.

The lightcurve and spectroscopy presented here offer a wealth of
information.  Much more detailed modeling will help in understanding the
nature of progenitor system, and could be a key to understanding important
aspects of the formation of Type~Ia supernovae. Better statistics
from larger unbiased nearby supernova searches will help greatly in
understanding the possibly multiple channels that produce SNe~Ia.

\acknowledgments

The authors are grateful to the technical and scientific staffs of the
University of Hawaii 2.2-meter telescope, the W.~M.~Keck Observatory,
Palomar Observatory, the QUEST-II collaboration, and HPWREN, for their
assistance in obtaining these data.  Some of the data presented herein
were obtained at the W. M. Keck Observatory, which is operated as a
scientific partnership among the California Institute of Technology,
the University of California, and the National Aeronautics and Space
Administration. The Observatory was made possible by the generous
financial support of the W. M. Keck Foundation.  We thank K~Barbary
and K.~Dawson for their assistance with the Keck LRIS observations,
and S.~Ferrell for assistance with photometry of the NEAT images.
The authors wish to recognize and acknowledge the very significant
cultural role and reverence that the summit of Mauna Kea has always
had within the indigenous Hawaiian community.  We are most fortunate to
have the opportunity to conduct observations from this mountain. This
work was supported in part by the Director, Office of Science, Office
of High Energy and Nuclear Physics, of the U.S. Department of Energy
under Contracts No. DE-FG02-92ER40704, by a grant from the Gordon \&
Betty Moore Foundation, by  National Science Foundation Grant Number
AST-0407297, and in France by support from CNRS/IN2P3, CNRS/INSU and PNC.
This research used resources of the National Energy Research Scientific
Computing Center, which is supported by the Office of Science of the U.S.
Department of Energy under Contract No. DE-AC03-76SF00098. We thank them
for a generous allocation of storage and computing time.  HPWREN is
funded by National Science Foundation Grant Number ANI-0087344, and
the University of California, San Diego.  Funding for the Sloan Digital
Sky Survey (SDSS) has been provided by the Alfred P. Sloan Foundation,
the Participating Institutions, the National Aeronautics and Space
Administration, the National Science Foundation, the U.S. Department
of Energy, the Japanese Monbukagakusho, and the Max Planck Society. The
SDSS Web site is http://www.sdss.org/.  GALEX is a NASA Small Explorer,
launched in April 2003. We gratefully acknowledge NASA's support for
construction, operation, and science analysis for the GALEX mission.

{\it Facilities:} 
\facility{PO:1.2m (QUEST-II)},
\facility{UH:2.2m (SNIFS)},
\facility{Keck:I (LRIS)},
\facility{GALEX}


\clearpage


\begin{table}
  \caption{SDSS \emph{i} Photometric Observations of \gj.}
  \vskip 10pt
  \centering
  \begin{tabular}{lllll}
    \tableline
    \tableline
    Julian Day & Camera & Filter & SDSS \emph{i} Magnitude \\
    \tableline
    2453636.91 & QUEST-II & RG610             & $< 20.15$         \\
    2453642.95 & QUEST-II & RG610             & $18.02 \pm 0.06$  \\
    2453647.09 & SNIFS    & Synthesized       & $17.78 \pm 0.10$  \\
    2453647.95 & QUEST-II & RG610             & $17.56 \pm 0.05$  \\
    2453648.91 & QUEST-II & Johnson \emph{I}  & $17.53 \pm 0.04$  \\
    2453651.78 & QUEST-II & Johnson \emph{I}  & $17.29 \pm 0.04$  \\
    2453654.91 & QUEST-II & Gunn \emph{i}     & $17.34 \pm 0.04$  \\
    2453656.91 & QUEST-II & Gunn \emph{i}     & $17.18 \pm 0.05$  \\
    2453667.84 & QUEST-II & RG610             & $16.98 \pm 0.06$  \\
    2453670.90 & QUEST-II & RG610             & $16.97 \pm 0.07$  \\
    2453670.90 & QUEST-II & RG610             & $17.00 \pm 0.06$  \\
    2453692.90 & QUEST-II & RG610             & $16.88 \pm 0.09$  \\
    2453692.94 & QUEST-II & RG610             & $17.24^{+0.36}_{-0.27}$ \\
    2453699.96 & SNIFS    & Synthesized       & $17.07  \pm 0.04$ \\
    2453702.70 & QUEST-II & RG610             & $17.03 \pm 0.06$  \\
    2453706.96 & SNIFS    & Synthesized       & $17.12  \pm 0.04$ \\
    2453706.98 & SNIFS    & Synthesized       & $17.07  \pm 0.04$ \\
    \tableline
  \end{tabular}
  \tablecomments{Synthesized photometry is from flux-calibrated SNIFS spectra}
  \label{tab:photometry}
\end{table}

\clearpage


\begin{table}
  \caption{\emph{V}-band photometry of \gj.}
  \vskip 10pt
  \centering
  \begin{tabular}{lllll}
    \tableline
    \tableline
    Julian Day & Camera & Filter & Johnson \emph{V} Magnitude \\
    \tableline
    2453642.95 & QUEST-II & RG610             & $17.90 \pm 0.09$ \\
    2453647.09 & SNIFS    & Synthesized       & $17.59 \pm 0.15$ \\
    2453647.95 & QUEST-II & RG610             & $17.57 \pm 0.10$ \\
    2453648.91 & QUEST-II & Johnson \emph{R}  & $17.48 \pm 0.04$ \\
    2453651.78 & QUEST-II & Johnson \emph{R}  & $17.34 \pm 0.05$ \\
    2453648.91 & QUEST-II & Johnson \emph{B}  & $17.48 \pm 0.08$ \\
    2453651.78 & QUEST-II & Johnson \emph{B}  & $17.27 \pm 0.08$ \\
    2453654.92 & QUEST-II & Gunn \emph{r}     & $17.37 \pm 0.04$ \\
    2453656.91 & QUEST-II & Gunn \emph{r}     & $17.35 \pm 0.04$ \\
    2453667.84 & QUEST-II & RG610             & $17.20 \pm 0.09$ \\
    2453670.90 & QUEST-II & RG610             & $17.22 \pm 0.10$ \\
    2453670.90 & QUEST-II & RG610             & $17.18 \pm 0.10$ \\
    2453699.96 & SNIFS    & Synthesized       & $17.41 \pm 0.15$ \\
    2453702.70 & QUEST-II & RG610             & $17.45 \pm 0.09$ \\
    2453706.96 & SNIFS    & Synthesized       & $17.46 \pm 0.15$ \\
    2453706.98 & SNIFS    & Synthesized       & $17.41 \pm 0.15$ \\
    \tableline
  \end{tabular}
  \tablecomments{QUEST-II photometry is based upon extrapolations from Johnson \emph{R},
                 Johnson \emph{B}, Gunn \emph{r}, and RG610 filters, using SNIFS spectra 
                 for the color corrections. Synthesized photometry is from flux-calibrated SNIFS
                 spectra.}
  \label{tab:photometry_V}
\end{table}

\clearpage


\begin{deluxetable}{lllcccll}
\tabletypesize{\scriptsize}
\tablecaption{Journal of Spectroscopic Observations of \gj.\label{tab:spectra}}
\tablewidth{0pt}
\tablehead{
\colhead{Julian Day}        &
\colhead{UTC Date}          &
\colhead{Range}             &
\colhead{Resolution\tablenotemark{a}}        &
\colhead{Exposure}          &
\colhead{Airmass}           &
\colhead{Facility}          &
\colhead{Conditions}        \\
\colhead{}                  &
\colhead{}                  &
\colhead{(\AA)}             &
\colhead{Blue : Red (\AA)}  &
\colhead{(s)}               &
\colhead{}                  &
\colhead{}                  &
\colhead{}      
}
\startdata
2453647.09 & 2005-10-03.59 & 3501-9994 & 2.6 : 3.5 & 1000 & 1.16 & UH88 + SNIFS & cloudy           \\
2453699.96 & 2005-11-25.46 & 3501-9994 & 2.6 : 3.5 & 1400 & 1.23 & UH88 + SNIFS & photometric      \\
2453706.83 & 2005-12-02.33 & 3225-9489 & 12  : 8.2 &  900 & 1.08 & Keck I + LRIS & non-photometric \\
2453706.96 & 2005-12-02.46 & 3501-9994 & 2.6 : 3.5 & 1400 & 1.31 & UH88 + SNIFS & near-photometric \\
2453706.98 & 2005-12-02.48 & 3501-9994 & 2.6 : 3.5 & 1400 & 1.46 & UH88 + SNIFS & near-photometric \\
2453768.23 & 2006-02-02.23 & 3500-6755 & 6.5 : 1.8 &  900 & 1.11 & Keck I + LRIS & photometric \\
\enddata
\tablenotetext{a}{\   LRIS resolution is seeing dependent}
\end{deluxetable}


\begin{deluxetable}{ccccccccccc}
\tabletypesize{\scriptsize}
\tablecaption{Emission line fits using two Gaussians plus continuum.
\label{tab:TwoGaussFits}}
\tablewidth{0pt}
\tablehead{
\colhead{Source}  &
\colhead{Line}    &
\colhead{Day}     &
\multicolumn{4}{c}{\hrulefill\hspace{0.05truein} Broad Component\hspace{0.05truein}\hrulefill}  &
\multicolumn{4}{c}{\hrulefill\hspace{0.05truein} Narrow Component\hspace{0.05truein}\hrulefill} \\
\colhead{}        & 
\colhead{}        &
\colhead{}        &
\colhead{$\lambda_{0}$} &
\colhead{$\sigma$} & 
\colhead{FWHM\tablenotemark{a}} & 
\colhead{Luminosity} &
\colhead{$\lambda_{0}$} &
\colhead{$\sigma$\tablenotemark{a}}      &
\colhead{FWHM\tablenotemark{b}} & 
\colhead{Luminosity} \\ 
\colhead{}              &
\colhead{}              &
\colhead{}              &
\colhead{(\AA)}         &
\colhead{(\AA)}         &
\colhead{(\kms)}        &
\colhead{(erg/s)}       &
\colhead{(\AA)}         &
\colhead{(\AA)}         &
\colhead{(\kms)}        &
\colhead{(erg/s)}       \\[-4.0ex]
}
\startdata
SNIFS& \Halpha\   & 11&6563.9&16.3&1670&$6.7\times10^{40}$&6563.3&3.5&unresolved        &$2.3\times 10^{40}$\\
SNIFS& \Halpha\   & 64&6564.8&20.7&2140&$7.7\times10^{40}$&6564.4&4.1&unresolved        &$3.6\times 10^{40}$\\
SNIFS& \Halpha\   & 71&6564.6&21.3&2210&$8.4\times10^{40}$&6564.5&4.0&unresolved        &$3.2\times 10^{40}$\\ 
LRIS & \Halpha    & 71&6564.5&18.1&1870&$5.4\times10^{40}$&6563.6&2.2&unresolved &$2.3\times 10^{40}$\\
LRIS & \Hbeta     & 71&4862.1&14.0&1710&$1.1\times10^{40}$&4863.4&3.6&unresolved &$6.9\times 10^{39}$\\
LRIS & \Hbeta     &133&4859.9&12.7&1780&$1.6\times10^{40}$&4862.4&1.4&unresolved        &$3.7\times 10^{39}$\\
LRIS & \ion{He}{1}&133&5876.8&20.7&1740&$1.2\times10^{40}$&5876.4&0.8&60    &$1.0\times 10^{39}$\\
\enddata
\tablenotetext{a}{Covariance between the width of the narrow component
and the broad component, or with the P~Cygni line, can lead to a
best-fit width that is slightly wider or narrower than the spectrograph
resolution.}
\tablenotetext{b}{FWHM is reported only when width due to spectrograph resolution can be meaningfully removed.}
\end{deluxetable}

\begin{deluxetable}{cccccccccccc}
\tabletypesize{\scriptsize}
\tablecaption{Emission line fits including P~Cygni profile
\label{tab:PCygniFits}}
\tablewidth{0pt}
\tablehead{
\colhead{Source}  &
\colhead{Line}    &
\colhead{Day}     &
\multicolumn{3}{c}{Broad Component}   &
\multicolumn{3}{c}{Narrow Component}  &
\multicolumn{3}{c}{P~Cygni Component} \\
\colhead{}        & 
\colhead{}        &
\colhead{}        &
\colhead{$\lambda_{0}$}         &
\colhead{$\sigma$}              & 
\colhead{FWHM\tablenotemark{a}} & 
\colhead{$\lambda_{0}$}         &
\colhead{$\sigma$}              &
\colhead{FWHM\tablenotemark{a}} & 
\colhead{$v_{phot}$\tablenotemark{b}} &
\colhead{$\tau_0$\tablenotemark{c}}   &
\colhead{$v_e$\tablenotemark{d}}     \\ 
\colhead{}              &
\colhead{}              &
\colhead{}              &
\colhead{(\AA)}         &
\colhead{(\AA)}         &
\colhead{(\kms)}        &
\colhead{(\AA)}         &
\colhead{(\AA)}         &
\colhead{(\kms)}        &
\colhead{(\kms)}        &
\colhead{}              &
\colhead{(\kms)}       \\[-4.0ex]
}
\startdata
LRIS & \Halpha    & 71&6562.1&19.5&2020&6563.7&2.3&unresolved&282&5.9&51\\
LRIS & \Hbeta     &133&4861.6&12.4&1740&4862.4&4.5&unresolved&251&189&6\\
LRIS & \ion{He}{1}&133&5880.2&27.6&3240&5876.1&1.0&unresolved& 60&0.3&50\\
\enddata
\tablenotetext{a}{FWHM is reported only when width due to spectrograph resolution can be meaningfully removed.}
\tablenotetext{b}{Photospheric velocity}
\tablenotetext{c}{Optical depth at the photosphere}
\tablenotetext{d}{e-folding velocity}
\end{deluxetable}

\clearpage

\begin{table}
  \caption{Dereddened magnitudes for the \gj\ host galaxy.}
  \vskip 10pt
  \centering
  \begin{tabular}{lccccc}
    \tableline
    \tableline
           &      & SDSS   & SDSS   & SDSS   & SDSS \\
    Source & Band & Parent & Child~A & Child~B & Red Template \\
    \tableline
    GALEX & FUV & \nodata            & \nodata          & \nodata          & \nodata \\
    GALEX & NUV & $20.74 \pm 0.26$   & \nodata          & \nodata          & \nodata \\
    SDSS  &  u  & $20.27 \pm 0.30$   & $20.76 \pm 0.38$ & $22.39 \pm 1.43$ & $+0.12$ \\
    SDSS  &  g  & $19.60 \pm 0.05$   & $20.13 \pm 0.06$ & $21.21 \pm 0.14$ & $+0.00$ \\
    SDSS  &  r  & $19.32 \pm 0.06$   & $19.95 \pm 0.08$ & $20.48 \pm 0.10$ & $-1.69$ \\
    SDSS  &  i  & $18.91 \pm 0.04$   & $19.97 \pm 0.10$ & $19.69 \pm 0.06$ & $-2.80$ \\
    SDSS  &  z  & $18.70 \pm 0.12$   & $19.09 \pm 0.13$ & $19.49 \pm 0.18$ & $-3.28$ \\
    \tableline
  \end{tabular}
  \tablecomments{SDSS dereddened magnitudes are given for the two deblended
  components as well as the SDSS parent object (\gj\ resides in the part
  of the galaxy characterized by component Child~A). The last column gives
  the $g-m_{\lambda}$ colors of the galaxy at $\alpha=45$:18:04.25, $\delta=-00$:33:30.7 (J2000)
  used as a template for the red component of the host of \gj. As this is 
  used as a template, we assign no uncertainty --- however doing so would make 
  our $\chi^2$/DOF far too small. The GALEX spatial resolution of 7\arcsec\ means that
  the host will be seen as a point source, so it cannot be decomposed into
  components matching the SDSS children. Note that when fitting for the age, we redden
  the model, rather than unredden the data in order to correctly account for
  differential extinction across filter bandpasses.}
  \label{tab:sdsshost}
\end{table}

\clearpage



\clearpage

\clearpage

\clearpage

\clearpage

\begin{figure}[htbp]
  \centering
  \includegraphics[width=1.2\textwidth, angle=90]{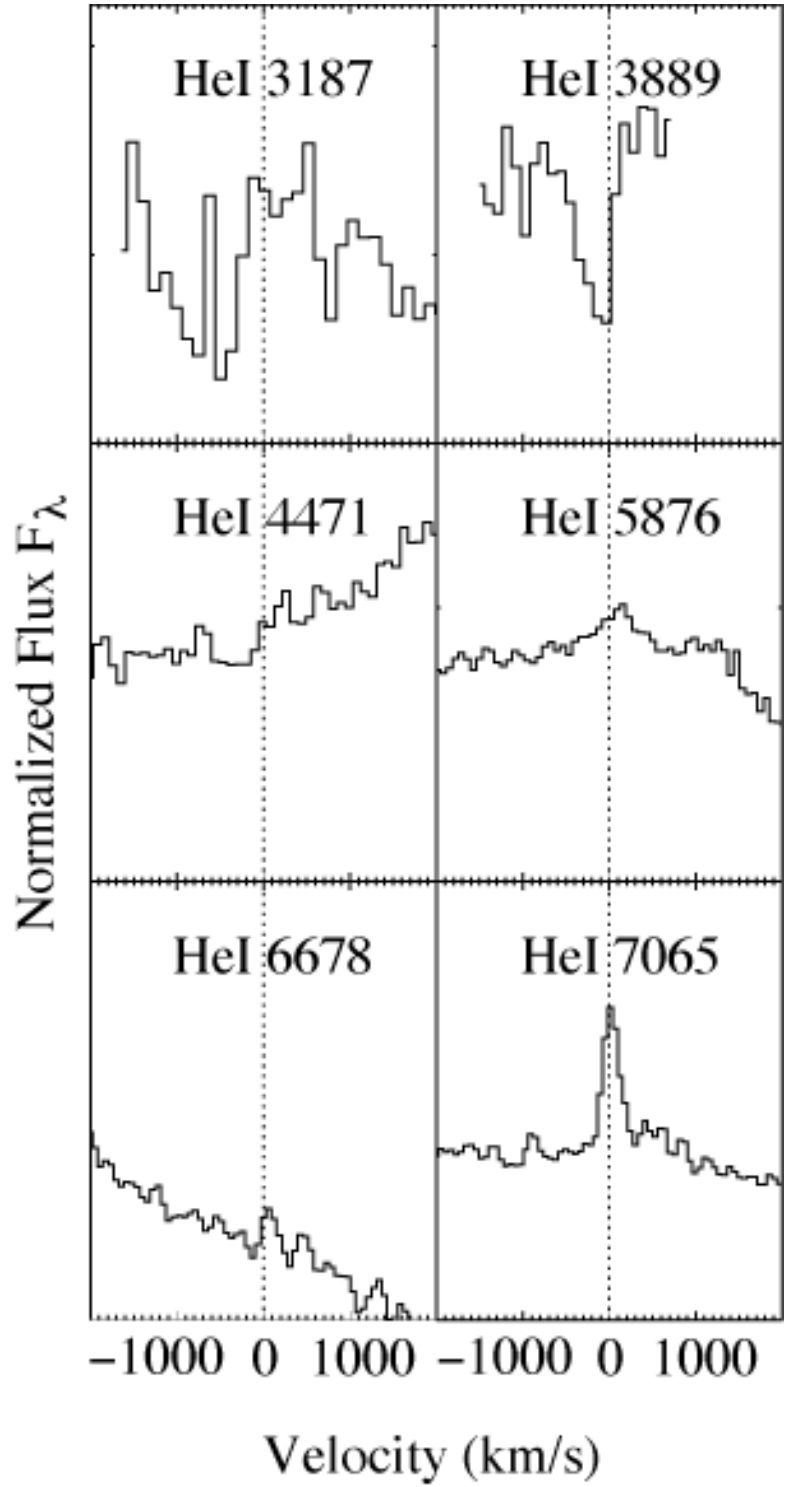}
  \caption{Line velocity profiles for \ion{He}{1} lines from our day~71 LRIS
    spectrum. Note that the feature at \ion{He}{1} 3889\AA\ is most likely
    to be the H8 line of the hydrogen Balmer series.}
 \label{fig:Hed336}
\end{figure}


\begin{figure}
 \centering
 \includegraphics[width=0.805\textwidth,clip=true]{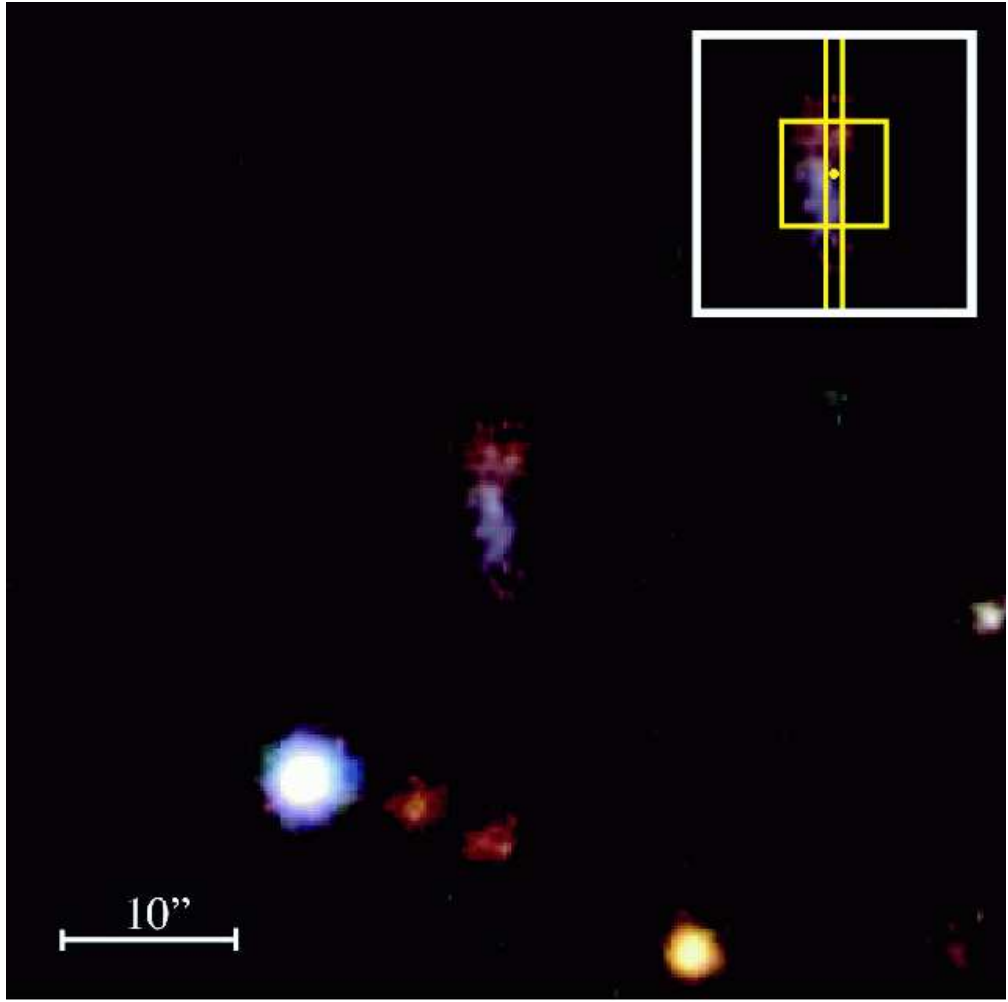}
 \caption{SDSS \emph{gri} color composite of the host galaxy of \gj\ with
 north up.  In the inset to the upper right, the SNIFS IFU 6x6 arcsecond
 field of view is shown as a square, the LRIS 1\arcsec\ wide slit runs
 north and south, and the location of \gj\ on the host is indicated by
 the dot.  The bar at the lower left is 10\arcsec.
 }
 \label{fig:sdsshost}
\end{figure}


\begin{figure}[htbp]
  \centering
  \includegraphics[width=0.75\textwidth,angle=90]{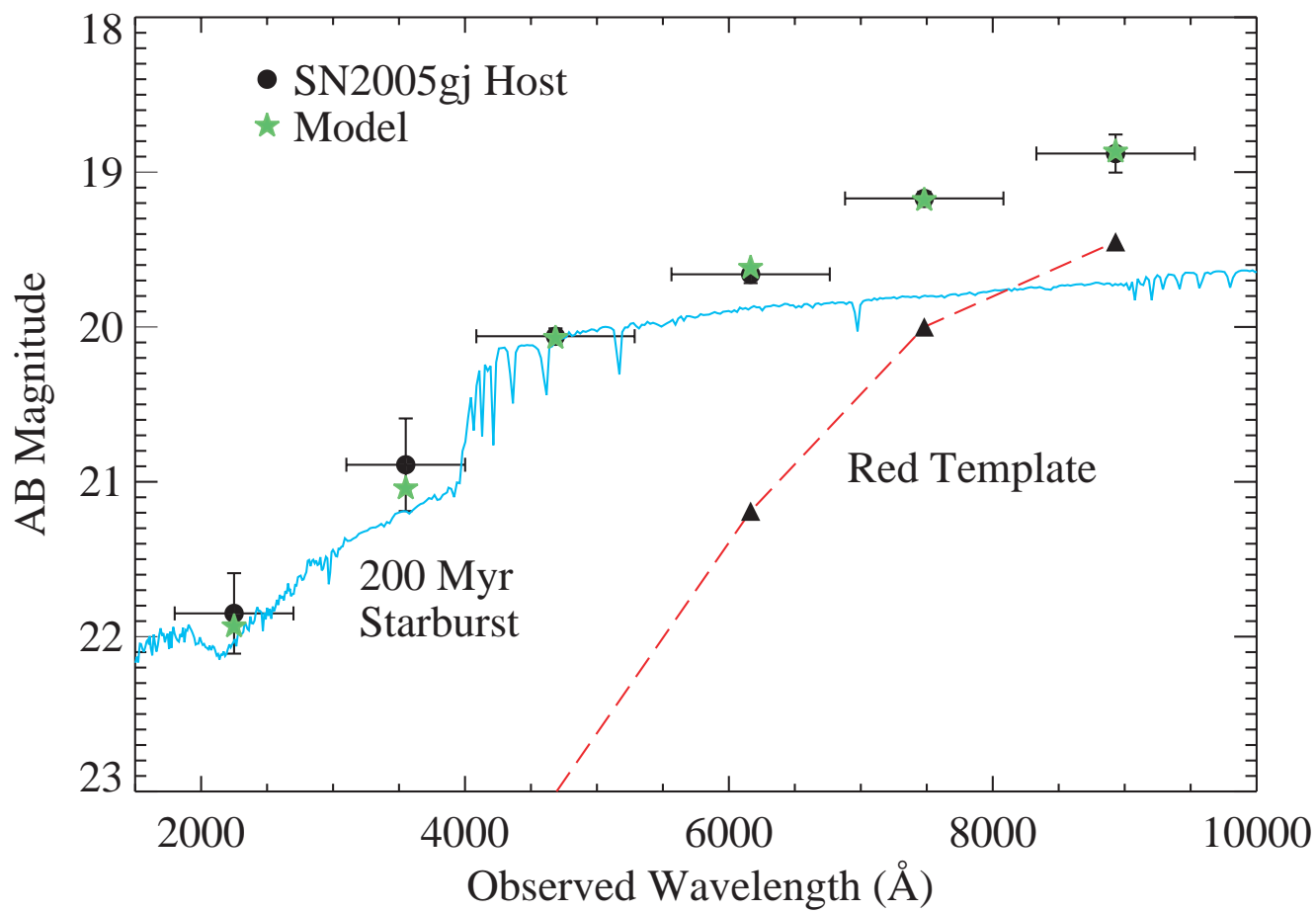}
  \caption{Spectrum of the host of \gj\ from UV and optical broadband
   photometry. Each black circle indicates the AB magnitude, uncorrected 
   for extinction, plotted at the mean wavelength, of each photometry
   point. The error bars in magnitude represent the photometric
   uncertainties, while the error bars in wavelength give an approximate
   indication of the wavelength range of each bandpass. Also shown are the
   components of a fit ---- the starburst models of \citet{leitherer99}
   plus a red galaxy template taken directly from a galaxy near the host
   of \gj.  In performing the fits, the starburst models were reddened
   and then integrated over the detailed filter bandpass. These starburst
   models were scaled and combined with the red template, and then the
   resulting model spectrum was scaled to fit the data. The results of
   the fit are indicated with green star symbols.}
  \label{fig:Host_fits}
\end{figure}

\end{document}